\documentclass[12pt,a4paper]{iopart}
\usepackage{iopams}
\usepackage{graphicx}
\usepackage{citesort}
\usepackage[normalem]{ulem}

\eqnobysec

\newcommand{\ee}{\mathrm{e}}
\newcommand{\ii}{\mathrm{i}}

\def\nn{\nonumber}

\def\diag{\mathrm{diag}}
\def\Vec#1{\mbox{\boldmath $#1$}}
\def\eqref#1{(\ref{#1})}
\newtheorem{thm}{Theorem}[section]
\newtheorem{prop}[thm]{Proposition}

\begin{document}

\title[Exact time-evolving scattering states with spin degrees of freedom]
{Exact time-evolving resonant states
for open double quantum-dot systems
with spin degrees of freedom}

\author{Akinori Nishino$^{1, 2}$ and Naomichi Hatano$^2$}

\address{$^1$ Faculty of Engineering, Kanagawa University, 
3-27-1 Rokkakubashi, Kanagawa-ku, Yokohama, Kanagawa, 221-8686, Japan}
\address{$^2$ Institute of Industrial Science, The University of Tokyo, 
5-1-5 Kashiwanoha, Kashiwa, Chiba 277-8574, Japan}
\ead{nishino@kanagawa-u.ac.jp}
\vspace{10pt}
\begin{indented}
\item[]October 2025
\end{indented}

\begin{abstract}
We study time-evolving resonant states in an open double quantum-dot system, 
taking into account spin degrees of freedom as well as both on-dot and interdot Coulomb interactions.
We exactly derived a non-Hermite effective Hamiltonian acting on the subspace of two quantum dots,
where the non-Hermiticity arises from an effect of infinite external leads connected to the quantum dots.
By diagonalizing the effective Hamiltonian, we identify four types of two-body resonant states.
For the initial states of localized two electrons with opposite spins on the quantum dots,
we exactly solve the time-dependent Schr\"odinger equation and obtain time-evolving two-body resonant states.
The time-evolving resonant states are normalizable since their wave function grows exponentially 
only inside a finite space interval that expands in time with electron velocity.
By using the exact solution, we analyze the survival and transition probabilities 
of localized two electrons on the quantum dots.
\end{abstract}

%
\vspace{2pc}
\noindent{\it Keywords}: open quantum systems, quantum dots, 
initial-value problems, resonant states, exceptional points, exact solutions, survival probability

\submitto{\jpa}
%
%
%


\section{Introduction}

Resonant states are quasi-stationary states with resonance energies that have a negative imaginary part.
They were originally introduced for the study of decaying states of unstable nuclei~\cite{Gamow_28ZPhysA}.
It is significant that resonant states are defined as a solution of 
the Schr\"odinger equation under the boundary conditions of purely outgoing wave
in open quantum systems, which is referred to as the Siegert boundary condition~\cite{Gamow_28ZPhysA,Siegert_39PR}.
The wave function of resonant states decays exponentially in time 
due to the imaginary part of the resonance energy, while
it diverges exponentially in space due to the imaginary part of the associated wave number.
To deal with such spatially diverging wave functions, we would need non-standard normalization methods 
such as the introduction of convergence factors~\cite{Zeldovich_61JETP,Hokkyo_65PTP,Berggren_68NPA}
or the complex-scaling method~\cite{Aguilar-Combes_71CMP,Balslev-Combes_71CMP,Simon_72CMP,%
Reinhardt_82AnnuRevPhysChem,Moiseyev_98PR,Moiseyev_11NHQP}. 
Then the study of resonant states has shifted to the application of resonant states
to an expansion of physical quantities,
which is derived from a completeness relation involving an integral over continuous states
and a finite sum over discrete states including resonant states~\cite{Berggren_68NPA,%
Romo_68NPA,GarciaCalderon-Peierls_76NPA,Berrondo-GarciaCalderon_77LNC,Lind_93PRA}.
It is remarkable that the expansion of Green's functions reduces to a finite sum over discrete states
if one of their variables is restricted to the region of a finite-range potential~\cite{More_71PRA,%
More-Gerjuoy_73PR,GarciaCalderon_76NPA,GarciaCalderon_82LNC,GarciaCalderon-Rubio_86NPA}.

Recently, resonant states have attracted considerable interest 
in the theoretical study of electron transport in open quantum systems.
The above resonant-state expansion of Green's functions enables us to calculate
the expansion of the transmission probability of an electron through multi-barrier tunneling structures;
each resonant state corresponds to a resonant peak in the transmission probability~\cite{%
GarciaCalderon-Romo-Rubio_93PRB,Romo-GarciaCalderon_94PRB,GarciaCalderon-Romo-Rubio_97PRB}.
For tight-binding models of open quantum-dot systems~\cite{Hatano-Sasada-Nakamura-Petrosky_08PTP,%
Hatano-Ordonez_14JMP,Ordonez-Hatano_17JPA,Ordonez-Hatano_17Chaos,Hatano_21JPCS},
the resonant state is characterized as an eigenstate of 
a non-Hermite effective Hamiltonian acting on the subspace of quantum dots
that is derived by the Feshbach formalism~\cite{Feshbach_58ARNS,Feshbach_58AP,Feshbach_62AP}.
Furthermore, the completeness relation involving resonant states and anti-resonant states
leads to the analysis of the time evolution of the survival probability on a quantum dot
and of the existence probability on external leads~\cite{%
Hatano-Ordonez_14JMP,Ordonez-Hatano_17JPA,Ordonez-Hatano_17Chaos,Hatano_21JPCS}.

A remaining issue in the study of resonance states is the incorporation of interactions.
In the previous work~\cite{Nishino-Hatano_24JPA},
we extended the concepts of resonant states and the Siegert boundary condition 
to an interacting case for the first time.
For an open double quantum-dot system with an interdot Coulomb interaction,
we exactly solved the time-dependent Schr\"odinger equation 
for the initial states of plane waves on the external leads
or of localized electrons on the two quantum dots.
In the latter case, we discovered {\it time-evolving resonant states}.
An essential difference from the spatially diverging wave functions of known resonant states 
as a solution of the time-independent Schr\"odinger equation is 
that the wave functions of the time-evolving resonant states exhibit exponential growth 
only inside a finite space interval that expands in time with electron velocity.
Clearly, the time-evolving resonant states are normalizable~\cite{%
Hatano-Sasada-Nakamura-Petrosky_08PTP,Hatano-Kawamoto-Feinberg_09PJP}.
Through the discovery of time-evolving resonant states, 
the concept of resonant states has acquired a physical meaning in its own right.

In the present article, we investigate an open double quantum-dot system with spin degrees of freedom, 
in which we take into account both on-dot and interdot Coulomb repulsions for localized electrons 
on the quantum dots~\cite{Nishino-Imamura-Hatano_12JPC,Nishino-Hatano-Ordonez_16JPC}.
First, under the many-body extension of the Siegert boundary condition,
we exactly derived a non-Hermite effective Hamiltonian 
acting on the subspace of the two quantum dots.
The exact effective Hamiltonian is independent of the energy, 
which is due to linear dispersion relations of the system,
and defines one-body and two-body resonant states as its eigenstates.
A major difference from the previous spinless case~\cite{Nishino-Hatano_24JPA} 
lies in the dimensionality of the subspace that the effective Hamiltonian acts on; 
the subspace is one-dimensional for two spinless electrons,
whereas it becomes four-dimensional for two electrons with opposite spins.
By diagonalizing the effective Hamiltonian,
we obtain four distinct types of two-body resonance energies and corresponding two-body resonant states.
It should be noted that two of the four two-body resonance energies merge into one 
at an exceptional point of the non-Hermite effective Hamiltonian.

Second, we solve the time-dependent Schr\"odinger equation
for the initial state of localized two electrons on the two quantum dots
and obtain exact time-evolving two-body resonant states.
The time-evolving resonant states on the quantum dots are given by a superposition of
the four types of two-body resonant states.
The wave functions of the time-evolving two-body resonant states on the external leads
decay exponentially in time and grow exponentially within a finite space interval,
which is similar to the spinless case.
A difference from the spinless case is that, depending on the choice of initial states, 
interference between two-body resonant states can appear during the exponential decay of the wave function.
We show that such properties of the initial states are classified 
by $so(4)$-type Lie-algebraic structure of the effective Hamiltonian.

Third, by using the exact time-evolving two-body resonant states,
we exactly calculate the survival probability of initial states of localized two electrons on the quantum dots
and the transition probability from the initial states to other states.
The lifetime of the initial states is determined by
the imaginary parts of the two-body resonance energies.
Among the four initial states classified by the $so(4)$-type Lie algebra, 
the lifetimes of two are independent of the interactions,
which is similar to the case of two spinless electrons~\cite{Nishino-Hatano_24JPA},
while those of the other two depend on the difference of the on-dot and the interdot interactions.
The former two initial states decay to the external leads without transferring to other states on the quantum dots, 
whereas the latter two are partially transferred to each other during the decay to the leads.
We note that such a purely exponential behavior is 
due to unbounded linear dispersion relations of the leads;
if there were the lower limit of the dispersion relations,
we would have found a non-exponential decay~\cite{%
Khalfin_68PZETF,Chiu-Sudarshan-Misra_77PRD,Petrosky-Tasaki-Prigogine_91PhysicaA,Petrosky-Ordonez-Prigogine_01PRA,%
Garmon-Petrosky-Simine-Segal_13FortschrPhys,Chakraborty-Sensarma_18PRB,Garmon-Noba-Ordonez-Segal_19PRD,%
Taira-Hatano-Nishino_2024preprint}.
We remark that the non-exponential decay is experimentally observed in 
trapped ions and ultra-cold atoms~\cite{Itano-Heizen-Bollinger-Wineland_90PRA,Wilkinson_97Nature,Fischer-GutierrezMedina-Raizen_01PRL}.
The measurement of the survival probability of interacting electrons on quantum dots 
is also in progress~\cite{Fujisawa_06Science}.

The paper is organized as follows: 
In Section~\ref{sec:openDQD}, 
an open double quantum-dot system with spin degrees of freedom as well as
both on-dot and interdot Coulomb interactions is introduced. 
We impose the extended Siegert boundary conditions to the two-electron states
and derive non-Hermite effective Hamiltonians.
In Section~\ref{sec:time-evolving-resonant-states},
in a special case of the system parameters,
we exactly solve the time-dependent Schr\"odinger equation
under the initial condition of localized two electrons on the quantum dots
and obtain time-evolving two-body resonant states.
In Section~\ref{sec:survival-transition-prob}, 
the survival probability of initial states on the two quantum dots
and the transition probability from the initial states to other states on the quantum dots 
are explicitly calculated by using the exact solutions.
The initial states are classified in terms of irreducible representations of the $so(4)$-type Lie algebra.
Section~\ref{sec:concluding-remarks} is devoted to concluding remarks.
In \ref{sec:algebraic-structure}, we elucidate
the $so(4)$-type Lie algebraic structure of the non-Hermite effective Hamiltonian.

\section{Open double quantum-dot systems}
\label{sec:openDQD}

\subsection{Hamiltonian}

\begin{figure}[t]
\begin{center}
{
\begin{picture}(220,150)(0,0)
\put(0,-5){\includegraphics[width=220pt]{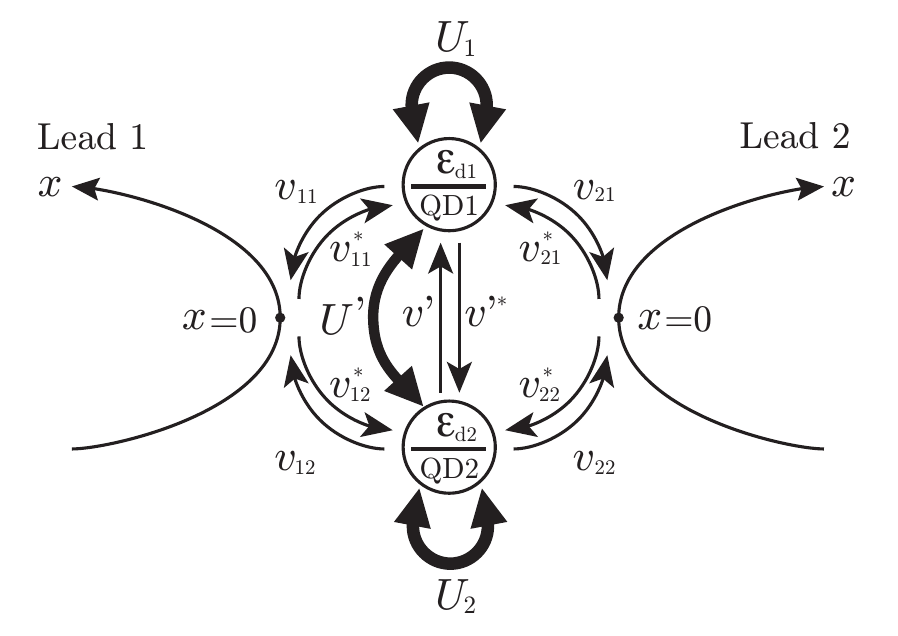}}
\end{picture}
}
\end{center}
\caption{\label{fig:GDQD-spin} A schematic diagram of the open double quantum-dot (QD) system.
The thick double-headed arrows represent on-dot and interdot Coulomb interactions.}
\end{figure}

We study quantum transport of electrons with spin degrees of freedom 
on an open double quantum-dot system~\cite{%
Nishino-Imamura-Hatano_12JPC,Nishino-Hatano-Ordonez_16JPC},
which consists of two quantum dots and two one-dimensional leads. 
As is illustrated in Figure~\ref{fig:GDQD-spin}, the $x$-axis is set along each lead;
the two quantum dots are connected to each other and to the origin $x=0$ of each lead.
We assume a linearized dispersion relation for each lead in the vicinity of the Fermi energy;
the positive Fermi velocity $v_{\rm F}>0$ corresponds to ``right-moving'' electrons
and the negative one $v_{\rm F}<0$ corresponds to ``left-moving'' electrons on each lead.

The Hamiltonian is given in the second-quantization form as
\begin{eqnarray}
\label{eq:open-DQD}
\fl
H=&\sum_{m=1,2}\sum_{\sigma=\uparrow, \downarrow}\int\! dx\,
   c^{\dagger}_{m\sigma}(x)
   v_{\rm F}\frac{1}{\ii}\frac{d}{dx}c_{m\sigma}(x)
 \nn\\
\fl
 &+\sum_{m=1,2}\sum_{\alpha=1,2}\sum_{\sigma=\uparrow, \downarrow}
   \big(v_{m\alpha}c^{\dagger}_{m\sigma}(0)d_{\alpha\sigma}
   +v^{\ast}_{m\alpha}d^{\dagger}_{\alpha\sigma}c_{m\sigma}(0)\big)
  +\sum_{\sigma=\uparrow, \downarrow}
   (v^{\prime}d^{\dagger}_{1\sigma}d_{2\sigma}+v^{\prime\ast}d^{\dagger}_{2\sigma}d_{1\sigma})
  \nn\\
\fl
 &+\sum_{\alpha=1,2}\sum_{\sigma=\uparrow, \downarrow}\epsilon_{{\rm d}\alpha}n_{\alpha\sigma}
  +\sum_{\alpha=1,2}U_{\alpha}n_{\alpha\uparrow}n_{\alpha\downarrow}
  +\sum_{\sigma, \tau=\uparrow, \downarrow}U^{\prime}n_{1\sigma}n_{2\tau},
\end{eqnarray}
where $c^{\dagger}_{m\sigma}(x)$ and $c_{m\sigma}(x)$ are
the creation- and annihilation-operators of electrons with spin $\sigma(=\uparrow, \downarrow)$
at position $x$ on the lead $m(=1, 2)$, respectively,
and $d^{\dagger}_{\alpha\sigma}$ and $d_{\alpha\sigma}$ are those on the quantum dot $\alpha(=1, 2)$, 
respectively.
We define a number operator $n_{\alpha\sigma}=d^{\dagger}_{\alpha\sigma}d_{\alpha\sigma}$
of electrons with spin $\sigma$ on the quantum dot $\alpha$.
Here and hereafter, we set $\hbar=1$.
The parameter $v_{m\alpha}$ is the transfer integral between the lead $m$ and the quantum dot $\alpha$,
$v^{\prime}$ is that between the two quantum dots with setting $v^{\prime}_{1}:=v^{\prime}$
and $v^{\prime}_{2}:=v^{\prime\ast}$ for notational convenience,
$\epsilon_{{\rm d}\alpha}$ is the energy level on the quantum dot $\alpha$,
$U_{\alpha}$ expresses the strength of the Coulomb repulsion on the quantum dot $\alpha$,
and $U^{\prime}$ expresses that of the interdot Coulomb repulsion.
We assume $U_{\alpha}\geq U^{\prime}$ due to physical requirements;
the case $U_{1}=U_{2}=U^{\prime}$ corresponds to a single quantum dot 
with two energy levels $\epsilon_{{\rm d}1}$ and $\epsilon_{{\rm d}2}$.
We remark that the system is regarded as a two-lead extension of 
the two-impurity Anderson model~\cite{Alexander-Anderson_64JPA}.
In a setup involving a few electrons, as will be discussed in the following sections, 
the Kondo effect intrinsic to the Anderson model does not appear. 
The Kondo effect would occur in many-body systems, not in few-body systems.

We note that, due to the linearized dispersion relations of the leads, 
the Hilbert space of right-moving electrons is decoupled from that of left-moving electrons.
Hence the Hamiltonian $H$ in Eq.~\eqref{eq:open-DQD} with a fixed value of the Fermi velocity $v_{\rm F}$
is Hermitian but not parity symmetric with respect to the space inversion $x\mapsto -x$ on each lead; 
the inversion maps the Hamiltonian for right-moving electrons to that for left-moving electrons.
It fact, the system is parity-time symmetric~\cite{Bender-Boettcher_98PRL}, that is,
symmetric under the combined transformation of 
the space inversion and time inversion.

\subsection{Time evolution of two-electron states}

We investigate time-evolving states of two electrons with opposite spins;
the cases of one electron and two spinless electrons were
studied in the previous work~\cite{Nishino-Hatano_24JPA}.
The time-evolving two-electron state is given in the form
\begin{eqnarray}
\label{eq:two-electron-state}
\fl
|\Psi_{\uparrow\downarrow}(t)\rangle
&=\Big(\sum_{m_{1}, m_{2}}\int\hspace{-2pt}dx_{1}dx_{2}\, 
  g_{m_{1}m_{2},\uparrow\downarrow}(x_{1},x_{2},t)
  c^{\dagger}_{m_{1}\uparrow}(x_{1})
  c^{\dagger}_{m_{2}\downarrow}(x_{2})
 \nn\\
\fl
&\quad
 +\sum_{m,\alpha, \sigma}\int\!\!dx\,
  e_{m\alpha,\sigma\overline{\sigma}}(x,t)
  c^{\dagger}_{m\sigma}(x)
  d^{\dagger}_{\alpha\overline{\sigma}}
 +\sum_{\alpha, \beta}f_{\alpha\beta,\uparrow\downarrow}(t)
  d^{\dagger}_{\alpha\uparrow}d^{\dagger}_{\beta\downarrow}
 \Big)|0\rangle.
\end{eqnarray}
Here $|0\rangle$ is the vacuum state, and we put
$\overline{\uparrow}=\downarrow$ and $\overline{\downarrow}=\uparrow$ 
in the sum on $\sigma$ of the second term.
The wave function $g_{m_{1}m_{2},\uparrow\downarrow}(x_{1},x_{2},t)$ describes 
two electrons on the two leads,
$e_{m\alpha,\sigma\overline{\sigma}}(x,t)$ 
describes two electrons, one of which is on the lead $m$ and the other is on the quantum dot $\alpha$,
and $f_{\alpha\beta,\uparrow\downarrow}(t)$ describes both of the two electrons 
on the quantum dots $\alpha$ and $\beta$.
One finds that the wave function $f_{\alpha\beta,\uparrow\downarrow}(t)$ with $\alpha=\beta$ corresponds to
double occupancy on one of the two quantum dots and $f_{\alpha\beta,\uparrow\downarrow}(t)$ 
with $\alpha\neq\beta$ corresponds to simultaneous occupancy on the two quantum dots.
For convenience, we also  define the wave functions $g_{m_{1}m_{2},\downarrow\uparrow}(x_{1},x_{2},t)$ and 
$f_{\alpha\beta,\downarrow\uparrow}(t)$ through the Fermionic anti-symmetry relations
\begin{eqnarray}
\label{eq:anti-sym_g-f}
&g_{m_{1}m_{2},\sigma\overline{\sigma}}(x_{1},x_{2},t)
 =-g_{m_{2}m_{1},\overline{\sigma}\sigma}(x_{2},x_{1},t),\quad
 f_{\alpha\beta,\sigma\overline{\sigma}}(t)
 =-f_{\beta\alpha,\overline{\sigma}\sigma}(t),
\end{eqnarray}
which are consistent with the anti-commutation relations among the creation-operators 
$c^{\dagger}_{m\sigma}(x)$ and $d^{\dagger}_{\alpha\sigma}$ in Eq.~\eqref{eq:two-electron-state}.

The time-dependent Schr\"odinger equation 
$\ii\partial_{t}|\Psi_{\uparrow\downarrow}(t)\rangle=H|\Psi_{\uparrow\downarrow}(t)\rangle$
with the Hamiltonian $H$ in Eq.~\eqref{eq:open-DQD}
gives the coupled differential equations for the wave functions
$g_{m_{1}m_{2},\sigma\overline{\sigma}}(x_{1},x_{2},t)$, 
$e_{m\alpha,\sigma\overline{\sigma}}(x,t)$ and $f_{\alpha\beta,\sigma\overline{\sigma}}(t)$
in Eq.~\eqref{eq:two-electron-state} as
\numparts
\label{eq:Sch-eq_2elec_1}
\begin{eqnarray}
\label{eq:Sch-eq_2elec-g_1}
\fl
&\ii\partial_{t}g_{m_{1}m_{2},\sigma\overline{\sigma}}(x_{1},x_{2},t)
  =v_{\rm F}\frac{1}{\ii}(\partial_{1}+\partial_{2})g_{m_{1}m_{2},\sigma\overline{\sigma}}(x_{1},x_{2},t)
  \nn\\
\fl
&+\sum_{\alpha}\big(v_{m_{2}\alpha}\delta(x_{2})e_{m_{1}\alpha,\sigma\overline{\sigma}}(x_{1},t)
  -v_{m_{1}\alpha}\delta(x_{1})e_{m_{2}\alpha,\overline{\sigma}\sigma}(x_{2},t)\big), 
  \\
\label{eq:Sch-eq_2elec-e_1}
\fl
&\ii\partial_{t}e_{m\alpha,\sigma\overline{\sigma}}(x,t)
 =\Big(v_{\rm F}\frac{1}{\ii}\partial_{x}+\epsilon_{{\rm d}\alpha}\Big)e_{m\alpha,\sigma\overline{\sigma}}(x,t)
 \nn\\
\fl
&+\sum_{n}v_{n\alpha}^{\ast}g_{mn,\sigma\overline{\sigma}}(x,0,t)
 +v^{\prime}_{\alpha}e_{m\overline{\alpha},\sigma\overline{\sigma}}(x,t)
 +\delta(x)\sum_{\beta}v_{m\beta}f_{\beta\alpha,\sigma\overline{\sigma}}(t),
 \\
\label{eq:Sch-eq_2elec-f_1}
\fl
&\ii\partial_{t}f_{\alpha\beta,\sigma\overline{\sigma}}(t)
  =(\epsilon_{{\rm d}\alpha}+\epsilon_{{\rm d}\beta}
   +U_{\alpha\beta})
   f_{\alpha\beta,\sigma\overline{\sigma}}(t)
  \nn\\
\fl
&+\sum_{m}\big(v^{\ast}_{m\alpha}e_{m\beta,\sigma\overline{\sigma}}(0,t)
  -v^{\ast}_{m\beta}e_{m\alpha,\overline{\sigma}\sigma}(0,t)\big)
  +v^{\prime}_{\alpha}f_{\overline{\alpha}\beta,\sigma\overline{\sigma}}(t)
  +v^{\prime}_{\beta}f_{\alpha\overline{\beta},\sigma\overline{\sigma}}(t).
\end{eqnarray}
\endnumparts
Here we put $\overline{\alpha}=3-\alpha$, $\partial_{t}=\partial/\partial t$,
$\partial_{i}=\partial/\partial x_{i}$ for $i=1,2$
and $U_{\alpha\beta}=\delta_{\alpha\beta}U_{\alpha}+(1-\delta_{\alpha\beta})U^{\prime}$
with Kronecker's delta $\delta_{\alpha\beta}$.

Similarly to the previous works~\cite{Nishino-Hatano_24JPA,Nishino-Imamura-Hatano_12JPC,%
Nishino-Hatano-Ordonez_16JPC},
we derive several relations among the wave functions 
from the set of Schr\"odinger equations~\eqref{eq:Sch-eq_2elec-g_1}, \eqref{eq:Sch-eq_2elec-e_1} 
and \eqref{eq:Sch-eq_2elec-f_1}.
The $\delta$-functions in Eq.~\eqref{eq:Sch-eq_2elec-g_1} indicate that 
the wave function $g_{m_{1}m_{2},\sigma\overline{\sigma}}(x_{1},x_{2},t)$
is discontinuous both at $x_{1}=0$ and $x_{2}=0$.
Since Eq.~\eqref{eq:Sch-eq_2elec-g_1} in each quadrant of the $(x_{1}, x_{2})$-plane
is equivalent to an advection equation,
the general solution $g_{m_{1}m_{2},\sigma\overline{\sigma}}(x_{1},x_{2},t)$
is given by an arbitrary function $F(x_{1}-v_{\rm F}t, x_{2}-v_{\rm F}t)$ 
of the two variables $x_{1}-v_{\rm F}t$ and $x_{2}-v_{\rm F}t$.
In other words, the wave function $g_{m_{1}m_{2},\sigma\overline{\sigma}}(x_{1},x_{2},t)$ 
has the following translation invariance:
\begin{eqnarray}
\label{eq:translation-invariance_g}
 g_{m_{1}m_{2},\sigma\overline{\sigma}}
 (x_{1}+v_{\rm F}\Delta t,x_{2}+v_{\rm F}\Delta t,t+\Delta t)
 =g_{m_{1}m_{2},\sigma\overline{\sigma}}(x_{1},x_{2},t)
\end{eqnarray}
if $x_{i}(x_{i}+v_{\rm F}\Delta t)>0$ for $i=1, 2$.
By integrating both sides of Eq.~\eqref{eq:Sch-eq_2elec-g_1} over 
the infinitesimal interval $0-<x_{1}<0+$ or $0-<x_{2}<0+$,
we obtain matching conditions of $g_{m_{1}m_{2},\sigma\overline{\sigma}}(x_{1},x_{2},t)$
at $x_{1}=0$ and $x_{2}=0$ as
\begin{eqnarray}
\label{eq:matching-cond_g}
\fl
&g_{m_{1}m_{2},\sigma\overline{\sigma}}(0+,x_{2},t)
 -g_{m_{1}m_{2},\sigma\overline{\sigma}}(0-,x_{2},t)
  -\frac{\ii}{v_{\rm F}}\sum_{\alpha}v_{m_{1}\alpha}e_{m_{2}\alpha,\overline{\sigma}\sigma}(x_{2},t)=0, 
  \nn\\
\fl
&g_{m_{1}m_{2},\sigma\overline{\sigma}}(x_{1},0+,t)
 -g_{m_{1}m_{2},\sigma\overline{\sigma}}(x_{1},0-,t)
  +\frac{\ii}{v_{\rm F}}\sum_{\alpha}v_{m_{2}\alpha}e_{m_{1}\alpha,\sigma\overline{\sigma}}(x_{1},t)=0.
\end{eqnarray}
Since the value 
$g_{mn,\sigma\overline{\sigma}}(x,0,t)$ at the discontinuous points,
which appears in Eq.~\eqref{eq:Sch-eq_2elec-e_1}, 
is not determined by the Schr\"odinger equations, 
we assume
\begin{eqnarray}
\label{eq:assumption_g}
&g_{m_{1}m_{2},\sigma\overline{\sigma}}(x_{1},0,t)
  =\frac{1}{2}\big(g_{m_{1}m_{2},\sigma\overline{\sigma}}(x_{1},0+,t)
     +g_{m_{1}m_{2},\sigma\overline{\sigma}}(x_{1},0-,t)\big).
\end{eqnarray}
By inserting the assumption in Eq.~\eqref{eq:assumption_g} 
into Eq.~\eqref{eq:Sch-eq_2elec-e_1} in the case $x\neq 0$
and using the second matching condition in Eqs.~\eqref{eq:matching-cond_g},
we obtain coupled differential equations for
$e_{m\alpha,\sigma\overline{\sigma}}(x,t)$ and $e_{m\overline{\alpha},\sigma\overline{\sigma}}(x,t)$
as
\begin{eqnarray}
\label{eq:Sch-eq_2elec-e_2}
&\ii\big(\partial_{t}+v_{\rm F}\partial_{x}\big)e_{m\alpha,\sigma\overline{\sigma}}(x,t)
  \nn\\
&=\Big(\epsilon_{{\rm d}\alpha}\mp\ii\frac{\Gamma_{\alpha\alpha}}{v_{\rm F}}\Big)
  e_{m\alpha,\sigma\overline{\sigma}}(x,t)
 +\Big(v^{\prime}_{\alpha}\mp\ii\frac{\Gamma_{\alpha\overline{\alpha}}}{v_{\rm F}}\Big)
  e_{m\overline{\alpha},\sigma\overline{\sigma}}(x,t)
 \nn\\
&\quad
 +\sum_{n}v_{n\alpha}^{\ast}g_{mn,\sigma\overline{\sigma}}(x,0\mp,t),
\end{eqnarray}
where we have introduced the band width 
$\Gamma_{\alpha\beta}=\sum_{m}v^{\ast}_{m\alpha}v_{m\beta}/2$ for $\alpha, \beta=1, 2$,
which turns out to play the role of the complex self-energy 
in Section~\ref{sec:time-evolving-2-body-resonant-states}.
Here we have obtained the relation in Eq.~\eqref{eq:Sch-eq_2elec-e_2} with the upper sign  
by eliminating $g_{mn,\sigma\overline{\sigma}}(x,0+,t)$
in the use of the matching condition in Eqs.~\eqref{eq:matching-cond_g}
and the relation with the lower sign by eliminating $g_{mn,\sigma\overline{\sigma}}(x,0-,t)$.
We note that the inhomogeneous term $g_{mn,\sigma\overline{\sigma}}(x,0-,t)$ 
in Eq.~\eqref{eq:Sch-eq_2elec-e_2}
describes two electrons, at least one of which is in the region $x<0$ of the leads,
and $g_{mn,\sigma\overline{\sigma}}(x,0+,t)$
describes two electrons, at least one of which is in the region $x>0$ of the leads.
Hence $g_{mn,\sigma\overline{\sigma}}(x,0-,t)$ includes incoming waves
for right-moving electrons and outgoing waves for left-moving electrons,
while $g_{mn,\sigma\overline{\sigma}}(x,0+,t)$ includes outgoing waves
for right-moving electrons and incoming waves for left-moving electrons.

\begin{figure}[t]
\begin{center}
{
\begin{picture}(220,100)(0,0)
\put(0,0){\includegraphics[width=220pt]{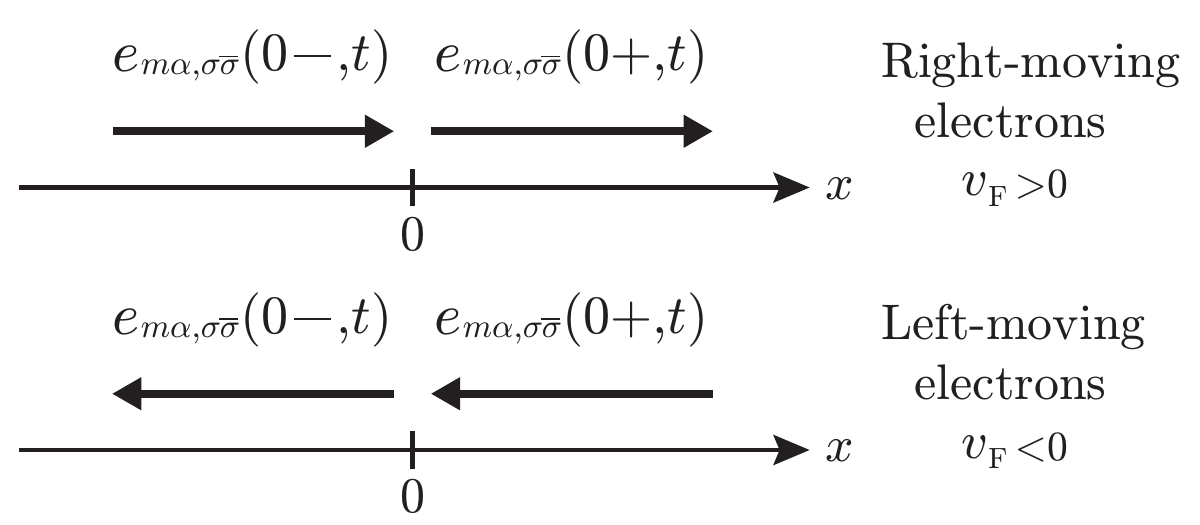}}
\end{picture}
}
\end{center}
\caption{\label{fig:GDQD-e(x)} Electron scattering around the origin $x=0$ of each lead described by
the wave functions $e_{m\alpha,\sigma\overline{\sigma}}(x,t)$.}
\end{figure}

Similarly to the above, Dirac's delta function in Eq.~\eqref{eq:Sch-eq_2elec-e_1} indicates that
the wave function $e_{m\alpha,\sigma\overline{\sigma}}(x,t)$ is discontinuous at $x=0$.
By integrating both sides of Eq.~\eqref{eq:Sch-eq_2elec-e_1} over the infinitesimal interval $0-<x<0+$,
we obtain a matching condition of the wave function $e_{m\alpha,\sigma\overline{\sigma}}(x,t)$
at the discontinuous point $x=0$ as
\begin{eqnarray}
\label{eq:matching-cond_e}
&e_{m\alpha,\sigma\overline{\sigma}}(0+,t)
 -e_{m\alpha,\sigma\overline{\sigma}}(0-,t)
 +\frac{\ii}{v_{\rm F}}\sum_{\beta}v_{m\beta}f_{\beta\alpha,\sigma\overline{\sigma}}(t)=0.
\end{eqnarray}
Similarly to Eq.~\eqref{eq:assumption_g}, we assume the value at the discontinuous point $x=0$ as
\begin{eqnarray}
\label{eq:assumption_e}
&e_{m\alpha,\sigma\overline{\sigma}}(0+,t)
 =\frac{1}{2}\big(e_{m\alpha,\sigma\overline{\sigma}}(0+,t)
 +e_{m\alpha,\sigma\overline{\sigma}}(0-,t)\big).
\end{eqnarray}
By using the assumption in Eq.~\eqref{eq:assumption_e}
with the matching condition in Eq.~\eqref{eq:matching-cond_e},
the equation~\eqref{eq:Sch-eq_2elec-f_1} is transformed to
\begin{eqnarray}
\label{eq:Sch-eq_2elec-f_2}
\ii\partial_{t}f_{\alpha\beta,\sigma\overline{\sigma}}(t)=
&\Big(\epsilon_{{\rm d}\alpha}+\epsilon_{{\rm d}\beta}+U_{\alpha\beta}
 \mp\ii\frac{\Gamma_{\alpha\alpha}+\Gamma_{\beta\beta}}{v_{\rm F}}\Big)f_{\alpha\beta,\sigma\overline{\sigma}}(t)
 \nn\\
&+\Big(v^{\prime}_{\alpha}\mp\ii\frac{\Gamma_{\alpha\overline{\alpha}}}{v_{\rm F}}\Big)
  f_{\overline{\alpha}\beta,\sigma\overline{\sigma}}(t)
 +\Big(v^{\prime}_{\beta}\mp\ii\frac{\Gamma_{\beta\overline{\beta}}}{v_{\rm F}}\Big)
  f_{\alpha\overline{\beta},\sigma\overline{\sigma}}(t)
 \nn\\
&+\sum_{m}\big(v^{\ast}_{m\alpha}e_{m\beta,\sigma\overline{\sigma}}(0\mp,t)
 -v^{\ast}_{m\beta}e_{m\alpha,\overline{\sigma}\sigma}(0\mp,t)\big).
\end{eqnarray}
Here we have obtained the relation with upper sign by eliminating $e_{m\alpha,\overline{\sigma}\sigma}(0+,t)$
and the relation with lower sign by eliminating $e_{m\alpha,\overline{\sigma}\sigma}(0-,t)$.
The wave functions $e_{m\alpha,\sigma\overline{\sigma}}(0-,t)$ 
describe two electrons, one of which is on the quantum dot $\alpha$
and the other is in the region $x<0$ of the lead $m$,
and $e_{m\alpha,\sigma\overline{\sigma}}(0+,t)$ 
describe two electrons, one of which is on the quantum dot $\alpha$
and the other is in the region $x>0$ of the lead $m$.
Hence, as is shown in Figure~\ref{fig:GDQD-e(x)}, the wave function 
$e_{m\alpha,\sigma\overline{\sigma}}(0-,t)$ includes incoming waves
for right-moving electrons and outgoing waves for left-moving electrons,
while $e_{m\alpha,\sigma\overline{\sigma}}(0+,t)$ includes outgoing waves
for right-moving electrons and incoming waves for left-moving electrons.

\subsection{Siegert boundary conditions and an effective Hamiltonian}

We propose an extension of the Siegert boundary conditions 
for the wave functions $g_{m_{1}m_{2},\sigma\overline{\sigma}}(x_{1},x_{2},t)$ 
and $e_{m\alpha,\sigma\overline{\sigma}}(x,t)$ of the two-electron states in Eq.~\eqref{eq:two-electron-state}.
The original Siegert boundary condition imposes a purely outgoing-wave condition 
on a single-electron wave function in order to characterize one-body
resonant states~\cite{Gamow_28ZPhysA,Siegert_39PR}.
Similarly, one-body anti-resonant states 
are obtained by imposing a purely incoming-wave condition on the wave function.

\begin{figure}[t]
\begin{center}
{
\begin{picture}(320,165)(0,0)
\put(0,-5){\includegraphics[width=320pt]{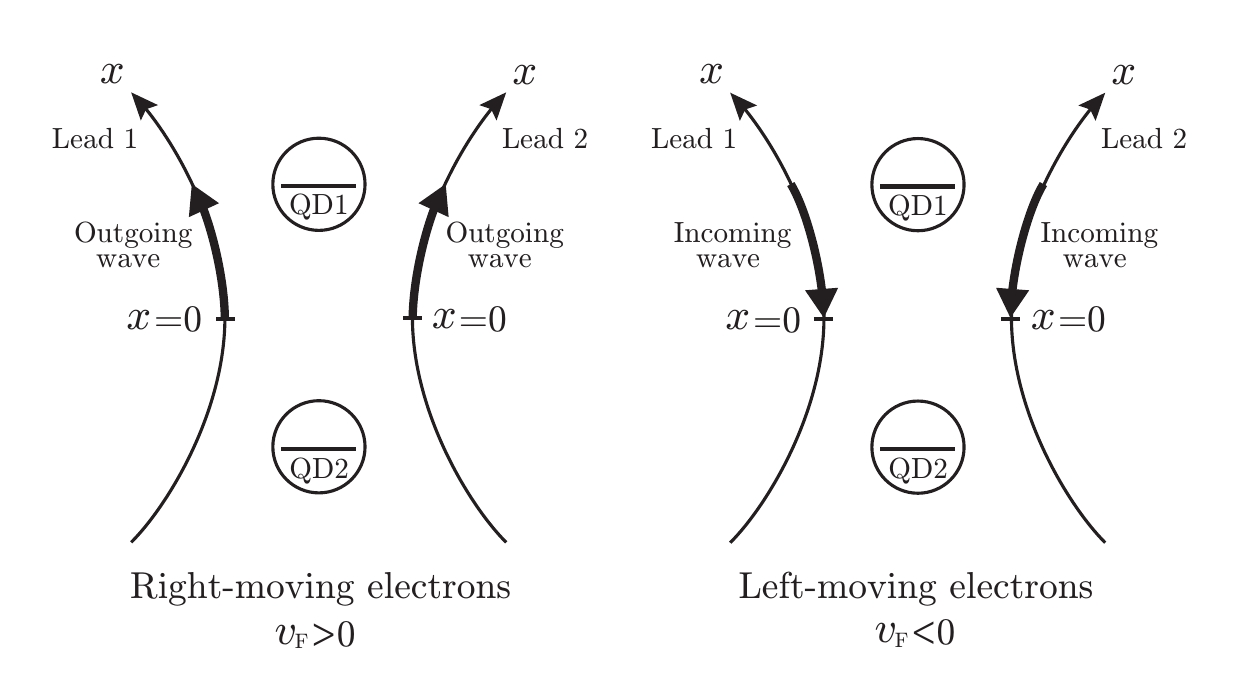}}
\end{picture}
}
\end{center}
\caption{\label{fig:GDQD-Siegert} 
Electron transport on the external leads of the open double quantum-dot system
under the Siegert boundary conditions in Eqs.~\eqref{eq:Siegert-2elec-bc_1}.
Outgoing waves from the origin appear for right-moving electrons with $v_{\rm F}>0$, 
while incoming waves toward the origin appear for left-moving electrons with $v_{\rm F}<0$. }
\end{figure}

In the present case of interacting two electrons, we consider boundary conditions
that there is no electron in the region $x<0$ of the leads as
\begin{eqnarray}
\label{eq:Siegert-2elec-bc_1}
&g_{m_{1}m_{2},\sigma\overline{\sigma}}(x_{1},x_{2},t)=0\quad
 \mbox{for } x_{1}<0 \mbox{ or } x_{2}<0,
 \nn\\
&e_{m\alpha,\sigma\overline{\sigma}}(x,t)=0\quad
 \mbox{for } x<0.
\end{eqnarray}
Under the conditions, purely outgoing waves appear for right-moving electrons,
while purely incoming waves appear for left-moving electrons,
as is illustrated in Fig.~\ref{fig:GDQD-Siegert}.
The former leads to two-body resonant states, 
while the latter leads to two-body anti-resonant states.
In the next section, we explicitly construct 
the two-body resonant states for right-moving electrons.

In a similar way, we consider boundary conditions
that there are no electrons in the region $x>0$ of the leads as
\begin{eqnarray}
\label{eq:Siegert-2elec-bc_2}
&g_{m_{1}m_{2},\sigma\overline{\sigma}}(x_{1},x_{2},t)=0\quad
 \mbox{for } x_{1}>0 \mbox{ or } x_{2}>0,
 \nn\\
&e_{m\alpha,\sigma\overline{\sigma}}(x,t)=0\quad
 \mbox{for } x>0.
\end{eqnarray}
Under the conditions, we obtain two-body anti-resonant states for right-moving electrons
and two-body resonant states for left-moving electrons.
The relations between the Siegert boundary conditions and
the two-body resonant/anti-resonant states are summarized in Table~\ref{tb:Siegert-ResonantState}.
It is clear that the boundary conditions in Eqs.~\eqref{eq:Siegert-2elec-bc_2}
are transformed to those in Eqs.~\eqref{eq:Siegert-2elec-bc_1} through the space inversion $x\mapsto -x$.
In what follows, we restrict our study to the boundary conditions in Eqs.~\eqref{eq:Siegert-2elec-bc_1}.

\begin{table}
\caption{\label{tb:Siegert-ResonantState}
Relations between the Siegert boundary conditions and 
the two-body resonant/anti-resonant states.}
\begin{indented}
\item[]\begin{tabular}{@{}ccc}
\br
\begin{tabular}{c}
 Siegert \\
 boundary conditions 
\end{tabular}
& \begin{tabular}{c}
   Right-moving electrons \\ 
   $v_{\rm F}>0$
  \end{tabular}
& \begin{tabular}{c}
   Left-moving electrons \\ 
   $v_{\rm F}<0$ 
  \end{tabular} \\
\mr
\begin{tabular}{c}
 No electrons in $x<0$ \\
 Eqs.~\eqref{eq:Siegert-2elec-bc_1}
\end{tabular}
& \begin{tabular}{c}
   Resonant state \\
   (Purely outgoing waves)   
  \end{tabular}
& \begin{tabular}{c}
   Anti-resonant state \\
   (Purely incoming waves)   
  \end{tabular}
  \\[10pt]
\begin{tabular}{c}
 No electrons in $x>0$ \\
 Eqs.~\eqref{eq:Siegert-2elec-bc_2}
\end{tabular}
& \begin{tabular}{c}
   Anti-resonant state \\
   (Purely incoming waves)   
  \end{tabular}
& \begin{tabular}{c}
   Resonant state \\
   (Purely outgoing waves)   
  \end{tabular}
  \\
\br
\end{tabular}
\end{indented}
\end{table}

We derive an effective Hamiltonian describing the time evolution of two-electron states
under the extended Siegert boundary conditions in Eqs.~\eqref{eq:Siegert-2elec-bc_1}.
First, we consider the wave functions $e_{m\alpha,\sigma\overline{\sigma}}(x,t)$
in the case of purely outgoing waves for right-moving electrons
or purely incoming waves for left-moving electrons.
By applying the conditions in Eqs.~\eqref{eq:Siegert-2elec-bc_1}
to the differential equations~\eqref{eq:Sch-eq_2elec-e_2}
for $e_{m\alpha,\sigma\overline{\sigma}}(x,t)$ with the upper sign chosen, we have
\begin{eqnarray}
\label{eq:Sch-eq_2elec-e_Siegert}
&\ii(\partial_{t}+v_{\rm F}\partial_{x})
 \Bigg(
 \begin{array}{c}
  e_{m1,\sigma\overline{\sigma}}(x,t) \\
  e_{m2,\sigma\overline{\sigma}}(x,t)
 \end{array}
 \Bigg)
 =H^{(1)}
 \Bigg(
 \begin{array}{c}
  e_{m1,\sigma\overline{\sigma}}(x,t) \\
  e_{m2,\sigma\overline{\sigma}}(x,t)
 \end{array}
 \Bigg),
\end{eqnarray}
where the matrix $H^{(1)}$ is given by
\begin{eqnarray}
\label{eq:eff-Hamiltonian_1-elec_1}
&H^{(1)}=
 \Bigg(
 \begin{array}{cc}
  \epsilon_{{\rm d}1}-\ii\Gamma_{11}/v_{\rm F} & v^{\prime}-\ii\Gamma_{12}/v_{\rm F} \\
  v^{\prime\ast}-\ii\Gamma_{21}/v_{\rm F} & \epsilon_{{\rm d}2}-\ii\Gamma_{22}/v_{\rm F} \\
 \end{array}
 \Bigg).
\end{eqnarray}
We note that the matrix $H^{(1)}$ in the case $v_{\rm F}=1$ is equal to the effective Hamiltonian 
that characterizes the one-body resonant state in the one-electron case~\cite{Nishino-Hatano_24JPA}.
However, in the present two-electron case, the matrix $H^{(1)}$ is not a Hamiltonian 
since the partial derivative in $x$ exists in the left-hand side of Eq.~\eqref{eq:Sch-eq_2elec-e_Siegert}.

The matrix $H^{(1)}$ is not Hermitian and
the imaginary term $-\ii\Gamma_{\alpha\beta}/v_{\rm F}$ in each element of $H^{(1)}$
expresses an effect of the external leads connected to the two quantum dots.
The term is independent of the energy $E$,
which is due to the unbounded linear dispersion relations of the leads.
As we shall see in Section~\ref{sec:resonance-energy}, 
the eigenvectors of the matrix $H^{(1)}$ lead to the general solution of
the wave function $e_{m\alpha,\sigma\overline{\sigma}}(x,t)$ in the region $x<0$ or $x>0$.
In order to determine the wave function $e_{m\alpha,\sigma\overline{\sigma}}(x,t)$
in the entire region of $x$,
we need to employ the matching condition at $x=0$ in Eq.~\eqref{eq:matching-cond_e} that involves
the wave function $f_{\alpha\beta,\sigma\overline{\sigma}}(t)$.

Next, we consider the wave functions $f_{\alpha\beta,\sigma\overline{\sigma}}(t)$
in the case of purely outgoing waves for right-moving electrons
or purely incoming waves for left-moving electrons.
By applying the conditions in Eqs.~\eqref{eq:Siegert-2elec-bc_1}
to the differential equations \eqref{eq:Sch-eq_2elec-f_2} for $f_{\alpha\beta,\sigma\overline{\sigma}}(t)$
with the upper sign chosen,  we obtain
\begin{eqnarray}
\label{eq:Sch-eq_2elec-f_3}
&\ii\partial_{t}
\left(
\begin{array}{c}
 f_{11,\sigma\overline{\sigma}}(t) \\
 f_{12,\sigma\overline{\sigma}}(t) \\
 f_{21,\sigma\overline{\sigma}}(t) \\
 f_{22,\sigma\overline{\sigma}}(t)
\end{array}
 \right)
=
H^{(2)}_{\rm eff}
\left(
 \begin{array}{c}
 f_{11,\sigma\overline{\sigma}}(t) \\
 f_{12,\sigma\overline{\sigma}}(t) \\
 f_{21,\sigma\overline{\sigma}}(t) \\
 f_{22,\sigma\overline{\sigma}}(t)
\end{array}
 \right),
\end{eqnarray}
where $H^{(2)}_{\rm eff}$ is a non-Hermite matrix given by
\begin{eqnarray}
\label{eq:eff-Hamiltonian_2-elec_1}
\fl
&H^{(2)}_{\rm eff}=
 \left(
 \begin{array}{cccc}
 \!\!\! 2\epsilon_{{\rm d}1}\!+\! U_{1}\!-\! 2\ii\Gamma_{11}/v_{\rm F}\!\!\! 
 & v^{\prime}\!-\!\ii\Gamma_{12}/v_{\rm F} & v^{\prime}\!-\!\ii\Gamma_{12}/v_{\rm F} & 0 \\
 v^{\prime\ast}\!-\!\ii\Gamma_{21}/v_{\rm F}
 & \!\!\! 2\overline{\epsilon}_{{\rm d}}\!+\! U^{\prime}\!\!-\! 2\ii\overline{\Gamma}/v_{\rm F}\!\!\! 
 & 0 & v^{\prime}\!-\!\ii\Gamma_{12}/v_{\rm F} \\
 v^{\prime\ast}\!-\!\ii\Gamma_{21}/v_{\rm F} & 0
 & \!\!\! 2\overline{\epsilon}_{{\rm d}}\!+\! U^{\prime}\!\!-\! 2\ii\overline{\Gamma}/v_{\rm F}\!\!\! 
 & v^{\prime}\!-\!\ii\Gamma_{12}/v_{\rm F} \\
 0 & v^{\prime\ast}\!-\!\ii\Gamma_{21}/v_{\rm F} & v^{\prime\ast}\!-\!\ii\Gamma_{21}/v_{\rm F}
 & \!\!\! 2\epsilon_{{\rm d}2}\!+\! U_{2}\!-\! 2\ii\Gamma_{22}/v_{\rm F}\!\!\! 
 \end{array}
 \right).
\end{eqnarray}
Since Eq.~\eqref{eq:Sch-eq_2elec-f_3} 
is in the form of the time-dependent Schr\"odinger equation,
the matrix $H^{(2)}_{\rm eff}$ is considered to be an {\it effective Hamiltonian}
that characterizes two-body resonant states.
It should be noted that the effective Hamiltonian $H^{(2)}_{\rm eff}$ is exactly derived
without any approximation such as the Markovian approximation.
In contrast to the wave function $e_{m\alpha,\sigma\overline{\sigma}}(x,t)$,
the wave function $f_{\alpha\beta,\sigma\overline{\sigma}}(t)$
is determined solely by the effective Hamiltonian $H^{(2)}_{\rm eff}$
and its initial conditions. 

It is remarkable that both the one-electron effective Hamiltonian~\cite{Nishino-Hatano_24JPA},
which is equal to the matrix $H^{(1)}$ in Eq.~\eqref{eq:eff-Hamiltonian_1-elec_1},
and the two-electron effective Hamiltonian $H^{(2)}_{\rm eff}$ 
in Eq.~\eqref{eq:eff-Hamiltonian_2-elec_1} are represented in the second-quantization form as 
\begin{eqnarray}
\label{eq:eff-Hamiltonian}
H_{\rm eff}
&=\sum_{\alpha, \sigma}
  \Big[\Big(\epsilon_{{\rm d}\alpha}-\ii\frac{\Gamma_{\alpha\alpha}}{v_{\rm F}}\Big)n_{\alpha\sigma}
  +\Big(v^{\prime}_{\alpha}-\ii\frac{\Gamma_{\alpha\overline{\alpha}}}{v_{\rm F}}\Big)
  d^{\dagger}_{\alpha\sigma}d_{\overline{\alpha}\sigma}\Big]
 \nn\\
&\quad
  +\sum_{\alpha}U_{\alpha}n_{\alpha\uparrow}n_{\alpha\downarrow}
  +U^{\prime}\sum_{\sigma, \tau}n_{1\sigma}n_{2\tau},
\end{eqnarray}
where we remind the readers that $n_{\alpha\sigma}=d^{\dagger}_{\alpha\sigma}d_{\alpha\sigma}$.
Here the action of the electron operators $d^{\dagger}_{\alpha\sigma}$ and $d_{\alpha\sigma}$ 
is restricted to the subspace of the two quantum dots. 
We have verified that this representation is extended to the case of four electrons,
which is the maximum number of electrons that can be accommodated on the two quantum dot.
The imaginary coefficients $-\ii\Gamma_{\alpha\alpha}/v_{\rm F}$ that appear in the terms of energy levels
play a role of complex potentials that absorb electrons from the dots for $v_{\rm F}>0$ 
or emit electrons into the dots for $v_{\rm F}<0$.
This complex potential may be called the self-energy in literature.
It is also noteworthy that, even in the case $v^{\prime}=0$ of decoupled two quantum dots, 
electrons on one quantum dot are transferred to the other via the external leads 
due to the self-energy terms with $-\ii\Gamma_{\alpha\overline{\alpha}}/v_{\rm F}$.
The last two interaction terms in Eq.~\eqref{eq:eff-Hamiltonian}
are not affected by the self-energies.

\section{Time-evolving resonant states}
\label{sec:time-evolving-resonant-states}

\subsection{Resonance energies in a special case}
\label{sec:resonance-energy}

The purpose of the present section is an {\it exact} construction of time-evolving states
for the initial states of localized two electrons with opposite spins on the two quantum dots.
To establish this, we consider a special case of the system parameters as
$v_{m\alpha}=v$, 
$\epsilon_{{\rm d}\alpha}=\epsilon_{{\rm d}}$,
$v^{\prime}=v^{\prime\ast}$ and $U_{\alpha}=U$ for $m=1, 2$ and $\alpha=1, 2$,
in which the system is symmetric 
with respect to both the exchange of the two leads and that of the two quantum dots.
We obtain an analytic expression of resonance energies and resonant states 
by diagonalizing the matrix $H^{(1)}$ in Eq.~\eqref{eq:eff-Hamiltonian_1-elec_1}
and the non-Hermite effective Hamiltonian $H^{(2)}_{\rm eff}$ in Eq.~\eqref{eq:eff-Hamiltonian_2-elec_1}.
The band width $\Gamma_{\alpha\beta}$ becomes independent of $\alpha$ and $\beta$;
we put $\Gamma_{\alpha\beta}=|v|^{2}=\Gamma$ for $\alpha, \beta=1, 2$.
For simplicity, we set the Fermi velocity $v_{\rm F}=1$ for right-moving electrons
and $v_{\rm F}=-1$ for the left-moving electrons.
In what follows, we will describe only the case $v_{\rm F}=1$, since
the time-evolving states in the case $v_{\rm F}=-1$ are obtained from those in the case $v_{\rm F}=1$
by the space inversion $x\mapsto -x$.

In the special case of the system parameters, the set of time-dependent Schr\"odinger equations
in Eqs.~\eqref{eq:Sch-eq_2elec-g_1}, \eqref{eq:Sch-eq_2elec-e_1} and \eqref{eq:Sch-eq_2elec-f_1} 
is simplified as
\numparts
\begin{eqnarray}
\label{eq:Sch-eq_2elec-g_5}
&\ii\partial_{t}g_{m_{1}m_{2},\sigma\overline{\sigma}}(x_{1},x_{2},t)
  =\frac{1}{\ii}(\partial_{1}+\partial_{2})g_{m_{1}m_{2},\sigma\overline{\sigma}}(x_{1},x_{2},t)
  \nn\\
&+v\sum_{\alpha}\big(\delta(x_{2})e_{m_{1}\alpha,\sigma\overline{\sigma}}(x_{1},t)
  -\delta(x_{1})e_{m_{2}\alpha,\overline{\sigma}\sigma}(x_{2},t)\big), 
  \\
\label{eq:Sch-eq_2elec-e_5}
&\ii\partial_{t}e_{m\alpha,\sigma\overline{\sigma}}(x,t)
 =\Big(\frac{1}{\ii}\partial_{x}+\epsilon_{{\rm d}}\Big)e_{m\alpha,\sigma\overline{\sigma}}(x,t)
 \nn\\
\fl
&+v^{\ast}\sum_{n}g_{mn,\sigma\overline{\sigma}}(x,0,t)
 +v^{\prime}e_{m\overline{\alpha},\sigma\overline{\sigma}}(x,t)
 +v\delta(x)\sum_{\beta}f_{\beta\alpha,\sigma\overline{\sigma}}(t),
 \\
\label{eq:Sch-eq_2elec-f_5}
&\ii\partial_{t}f_{\alpha\beta,\sigma\overline{\sigma}}(t)
  =(2\epsilon_{{\rm d}}+U_{\alpha\beta})
   f_{\alpha\beta,\sigma\overline{\sigma}}(t)
  \nn\\
&+v^{\ast}\sum_{m}\big(e_{m\beta,\sigma\overline{\sigma}}(0,t)
  -e_{m\alpha,\overline{\sigma}\sigma}(0,t)\big)
  +v^{\prime}f_{\overline{\alpha}\beta,\sigma\overline{\sigma}}(t)
  +v^{\prime}f_{\alpha\overline{\beta},\sigma\overline{\sigma}}(t),
\end{eqnarray}
\endnumparts
where $U_{\alpha\beta}=\delta_{\alpha\beta}U+(1-\delta_{\alpha\beta})U^{\prime}$.
Following the discussion in the previous section, 
we construct a general solution for the wave functions 
$e_{m\alpha,\sigma\overline{\sigma}}(x,t)$ and $f_{\alpha\beta,\sigma\overline{\sigma}}(t)$.

First, the matrix $H^{(1)}$ in Eq.~\eqref{eq:eff-Hamiltonian_1-elec_1} 
for the wave function $e_{m\alpha,\sigma\overline{\sigma}}(x,t)$
becomes 
\begin{eqnarray}
\label{eq:eff-Hamiltonian_1-elec_2}
&H^{(1)}=
 \Bigg(
 \begin{array}{cc}
  \epsilon_{{\rm d}}-\ii\Gamma & v^{\prime}-\ii\Gamma \\
  v^{\prime}-\ii\Gamma & \epsilon_{{\rm d}}-\ii\Gamma \\
 \end{array}
 \Bigg)
\end{eqnarray}
for right-moving electrons with $v_{\rm F}=1$.
Through the similarity transformation with the orthogonal matrix
\begin{eqnarray}
&S_{0}=\frac{1}{\sqrt{2}}
 \Bigg(
 \begin{array}{cc}
  1 & 1 \\
  -1 & 1
 \end{array}
 \Bigg),
\end{eqnarray}
the matrix $H^{(1)}$ is diagonalized as 
\begin{eqnarray}
\label{eq:1-resonance-energy_2}
 S^{-1}_{0}H^{(1)}S_{0}
&=
 \Bigg(
 \begin{array}{cc}
  \epsilon_{{\rm d}}-v^{\prime} & 0 \\
  0 & \epsilon_{{\rm d}}+v^{\prime}-2\ii\Gamma \\
 \end{array}
 \Bigg)
 =:
 \Bigg(
 \begin{array}{cc}
  E^{(1)}_{{\rm R}+} & 0 \\[-2pt]
  0 & E^{(1)}_{{\rm R}-}
 \end{array}
 \Bigg),
\end{eqnarray}
which provides the general solution $e_{m\alpha,\sigma\overline{\sigma}}(x,t)$
of the partial differential equation~\eqref{eq:Sch-eq_2elec-e_Siegert} with $v_{\rm F}=1$ as
\begin{eqnarray}
\label{eq:general-solution_e}
\fl
&\Bigg(
 \begin{array}{c}
  e_{m1,\sigma\overline{\sigma}}(x,t) \\
  e_{m2,\sigma\overline{\sigma}}(x,t)
 \end{array}
 \Bigg)
 =S_{0}
 \Bigg(
 \begin{array}{c}
  \sqrt{2}D_{m,\sigma\overline{\sigma},+}(x-t)\ee^{-\ii E^{(1)}_{{\rm R}+}(x+t)/2} \\[-2pt]
  \sqrt{2}D_{m,\sigma\overline{\sigma},-}(x-t)\ee^{-\ii E^{(1)}_{{\rm R}-}(x+t)/2}
 \end{array}
 \Bigg)
 \nn\\
\fl
&=
 \Bigg(
 \begin{array}{c}
   D_{m,\sigma\overline{\sigma},+}(x-t)\ee^{-\ii E^{(1)}_{{\rm R}+}(x+t)/2}
  +D_{m,\sigma\overline{\sigma},-}(x-t)\ee^{-\ii E^{(1)}_{{\rm R}-}(x+t)/2} \\
  -D_{m,\sigma\overline{\sigma},+}(x-t)\ee^{-\ii E^{(1)}_{{\rm R}+}(x+t)/2}
  +D_{m,\sigma\overline{\sigma},-}(x-t)\ee^{-\ii E^{(1)}_{{\rm R}-}(x+t)/2}
 \end{array}
 \Bigg)
\end{eqnarray}
for $x<0$ or $x>0$.
Here $D_{m,\sigma\overline{\sigma},\pm}(x-t)$ is an arbitrary function of the variable $x-t$,
which shall be determined by the initial conditions of $e_{m\alpha,\sigma\overline{\sigma}}(x,t)$
and the matching condition in Eq.~\eqref{eq:matching-cond_e}.
The solution~\eqref{eq:general-solution_e} indicates that
the real eigenvalue $E^{(1)}_{{\rm R}+}$ corresponds to a steady state which
consists of a bound state of an electron on the quantum dots
and a scattering state of the other electron on the entire leads,
while the complex eigenvalue $E^{(1)}_{{\rm R}-}$ with a negative imaginary part 
corresponds to a one-body resonant state with the other electron on the leads.

The emergence of the steady state with the real eigenvalue $E^{(1)}_{{\rm R}+}$
is understood by the separation of variables
for the coupled differential equations~\eqref{eq:Sch-eq_2elec-e_5}.
In fact, the wave function
\begin{eqnarray}
 e^{({\rm odd})}_{m,\sigma\overline{\sigma}}(x,t)
 =\frac{1}{\sqrt{2}}\big(e_{m1,\sigma\overline{\sigma}}(x,t)-e_{m2,\sigma\overline{\sigma}}(x,t)\big)\quad
 \mbox{for } m=1, 2,
\end{eqnarray}
which is an odd function with respect to the exchange of the two quantum dots,
is decoupled from the wave functions $g_{mn,\sigma\overline{\sigma}}(x_{1},x_{2},t)$
for $m, n=1, 2$,
which describe the two electrons on the leads, since it satisfies the differential equation
\begin{eqnarray}
 &\big(\ii(\partial_{t}+\partial_{x})-E^{(1)}_{{\rm R}+}\big)e^{({\rm odd})}_{m,\sigma\overline{\sigma}}(x,t)=0
 \quad \mbox{for } x\neq 0
\end{eqnarray}
without imposing the Siegert boundary conditions in Eqs.~\eqref{eq:Siegert-2elec-bc_1}.
On the other hand, the even wave function
\begin{eqnarray}
 e^{({\rm even})}_{m,\sigma\overline{\sigma}}(x,t)
 =\frac{1}{\sqrt{2}}\big(e_{m1,\sigma\overline{\sigma}}(x,t)+e_{m2,\sigma\overline{\sigma}}(x,t)\big)\quad
 \mbox{for } m=1, 2
\end{eqnarray}
of the one-body resonant state with the resonance energy $E^{(1)}_{{\rm R}-}$
is coupled to the wave functions $g_{mn,\sigma\overline{\sigma}}(x_{1},x_{2},t)$.
Hence the electron on the quantum dots that is described by 
the wave function $e^{({\rm even})}_{m,\sigma\overline{\sigma}}(x,t)$ decays to the leads.

Next, the Hamiltonian matrix $H^{(2)}_{\rm eff}$ in Eq.~\eqref{eq:eff-Hamiltonian_2-elec_1} 
for the wave function $f_{\alpha\beta,\sigma\overline{\sigma}}(t)$ is simplified 
in the special case of the parameters as
\begin{eqnarray}
\label{eq:eff-Hamiltonian_2-elec_2}
\fl
&H^{(2)}_{\rm eff}=
 \left(
 \begin{array}{cccc}
 2\epsilon_{{\rm d}}+U-2\ii\Gamma\!\!
 & v^{\prime}-\ii\Gamma & v^{\prime}-\ii\Gamma & 0 \\
 v^{\prime}-\ii\Gamma
 & \!\! 2\epsilon_{{\rm d}}+U^{\prime}-2\ii\Gamma \!\!
 & 0 & v^{\prime}-\ii\Gamma \\
 v^{\prime}-\ii\Gamma & 0
 & \!\! 2\epsilon_{{\rm d}}+U^{\prime}-2\ii\Gamma \!\! 
 & v^{\prime}-\ii\Gamma \\
 0 & v^{\prime}-\ii\Gamma & v^{\prime}-\ii\Gamma
 & \!\! 2\epsilon_{{\rm d}}+U-2\ii\Gamma
 \end{array}
 \right)
\end{eqnarray}
for right-moving electrons.
As we shall see below, the Hamiltonian matrix $H^{(2)}_{\rm eff}$ has an exceptional point
at the system parameters satisfying $U-U^{\prime}=4\Gamma$ with $v^{\prime}=0$.

By using the orthogonal matrix defined by
\begin{eqnarray}
\label{eq:orthogonal-matrix_1}
 S_{1}=\frac{1}{\sqrt{2}}
 \left(
 \begin{array}{cccc}
   1 & 0  & 0 & 1 \\
   0 & 1 & 1 & 0 \\
   0 & -1 & 1 & 0 \\
   -1 & 0 & 0 & 1
 \end{array}
 \right),
\end{eqnarray}
the Hamiltonian matrix $H^{(2)}_{\rm eff}$ is block-diagonalized as in
\begin{eqnarray}
\label{eq:eff-Hamiltonian_2-elec_3}
\fl
 S_{1}^{-1}H^{(2)}_{\rm eff}S_{1}=
 \left(
 \begin{array}{cccc}
   2\epsilon_{\rm d}\!+\! U-\! 2\ii\Gamma & 0 & 0 & 0 \\
   0 & 2\epsilon_{\rm d}\!+\! U^{\prime}\!-\! 2\ii\Gamma & 0 & 0 \\
   0 & 0 & 2\epsilon_{\rm d}\!+\! U^{\prime}\!-\! 2\ii\Gamma & 2(v^{\prime}\!-\!\ii\Gamma) \\
   0 & 0 & 2(v^{\prime}\!-\!\ii\Gamma) & 2\epsilon_{\rm d}\!+\! U\!-\! 2\ii\Gamma
 \end{array}
 \right).
\end{eqnarray}
The mathematical meaning of the orthogonal matrix $S_{1}$ 
in Eq.~\eqref{eq:orthogonal-matrix_1} is understood by an algebraic structure 
of the effective Hamiltonian $H_{\rm eff}$ in Eq.~\eqref{eq:eff-Hamiltonian},
which shall be described in \ref{sec:algebraic-structure}.
Furthermore, in order to diagonalize the remaining $2\times 2$ block, we introduce the matrix
\begin{eqnarray}
\label{eq:S-matrix_2}
&S_{2}=
 \left(
 \begin{array}{cccc}
   1 & 0 & 0 & 0 \\
   0 & 1 & 0 & 0 \\
   0 & 0 & p_{+} & p_{-} \\
   0 & 0 & q_{+} & q_{-}
 \end{array}
 \right),
\end{eqnarray}
where the matrix elements $p_{\pm}$ and $q_{\pm}$ are characterized by the relations
\begin{eqnarray}
\label{eq:pq_def}
&\frac{p_{\pm}}{q_{\pm}}
  =\frac{4(v^{\prime}-\ii\Gamma)}{\Delta U+\xi_{\pm}}
  =\frac{-\Delta U+\xi_{\pm}}{4(v^{\prime}-\ii\Gamma)},\quad
 p_{+}q_{-}-p_{-}q_{+}=1
\end{eqnarray}
with $\overline{U}=(U+U^{\prime})/2$, $\Delta U=U-U^{\prime}$ and
\begin{eqnarray}
\label{eq:xi_def}
&\xi_{\pm}=\pm\xi=\pm\sqrt{(\Delta U)^{2}+16(v^{\prime}-\ii\Gamma)^{2}}.
\end{eqnarray}
We note that the matrix $S_{2}$ in Eq.~\eqref{eq:S-matrix_2} is not unitary
since the Hamiltonian matrix $H^{(2)}_{\rm eff}$ is not Hermitian.
Through the similarity transformation with the matrix $S_{2}$ in Eq.~\eqref{eq:S-matrix_2},
the matrix in Eq.~\eqref{eq:eff-Hamiltonian_2-elec_3} is diagonalized as in
\begin{eqnarray}
\label{eq:2-resonance-energy}
\fl
&S_{2}^{-1}S_{1}^{-1}H^{(2)}_{\rm eff}S_{1}S_{2}
  \nn\\
\fl
&=\diag\Big(2\epsilon_{\rm d}+U-2\ii\Gamma,
 2\epsilon_{\rm d}+U^{\prime}-2\ii\Gamma,
 2\epsilon_{\rm d}+\overline{U}-2\ii\Gamma+\frac{\xi_{+}}{2},
 2\epsilon_{\rm d}+\overline{U}-2\ii\Gamma+\frac{\xi_{-}}{2}\Big)
 \nn\\
\fl
&=:\diag(E^{(2)}_{{\rm R}, 1}, E^{(2)}_{{\rm R}, 2}, E^{(2)}_{{\rm R}, 3+}, E^{(2)}_{{\rm R}, 3-}).
\end{eqnarray}
Thus we obtain four types of {\it two-body} resonance energies.

The four complex eigenvalues in Eq.~\eqref{eq:2-resonance-energy} 
are resonance energies giving two-body resonant states~\cite{Nishino-Hatano_24JPA}.
Here the resonance energies $E^{(2)}_{{\rm R}, 1}$ and $E^{(2)}_{{\rm R}, 2}$ share
the same imaginary part $-2\Gamma$, which is independent of the interactions. 
We note that the resonance energy $E^{(2)}_{{\rm R}, 2}$ appeared
in the case of two spinless electrons~\cite{Nishino-Hatano_24JPA},
which is understood by an algebraic structure 
of the effective Hamiltonian $H_{\rm eff}$ in Eq.~\eqref{eq:eff-Hamiltonian}
(see \ref{sec:algebraic-structure}).
It is remarkable that the imaginary parts of the resonance energies $E^{(2)}_{{\rm R}, 3\pm}$ depend on
the difference $\Delta U$ of the interaction parameters $U$ and $U^{\prime}$, 
which is an essential difference from the spinless case~\cite{Nishino-Hatano_24JPA}.

\begin{figure}[t]
\begin{center}
{
\begin{picture}(360,160)(0,0)
\put(0,-5){\includegraphics[width=360pt]{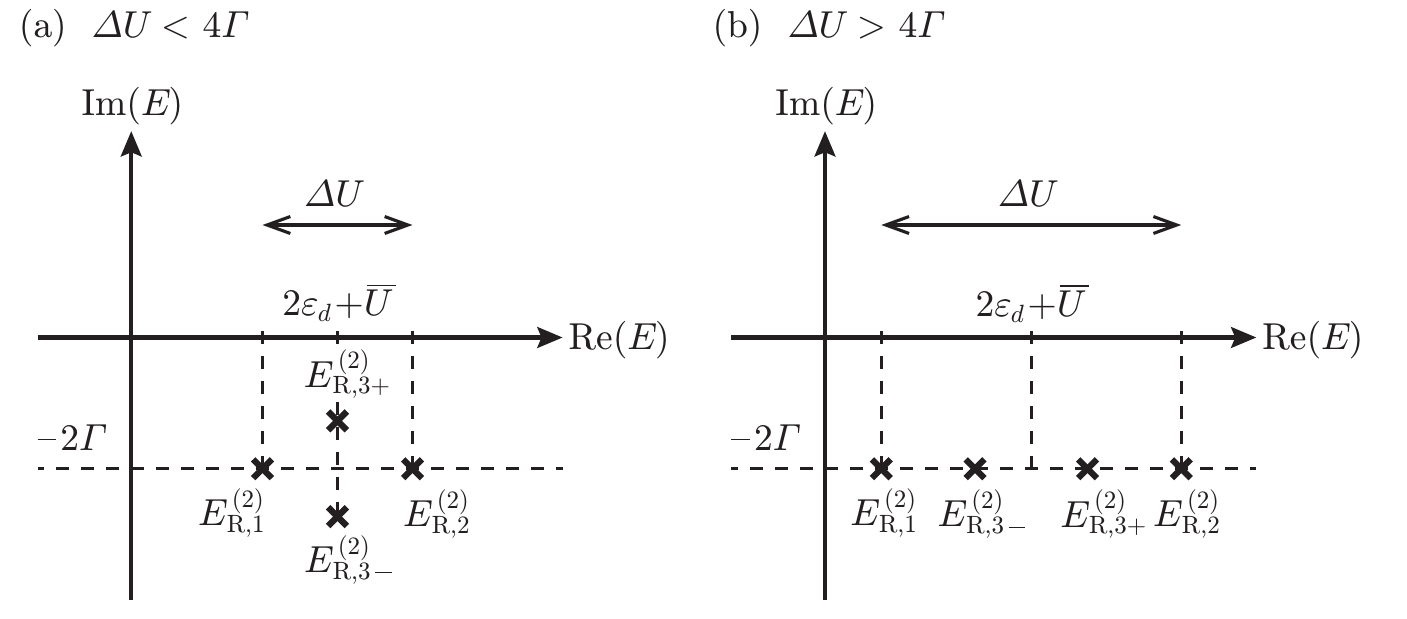}}
\end{picture}
}
\end{center}
\caption{\label{fig:resonance-energies}
Arrangement of the two-body resonance energies $E^{(2)}_{{\rm R}, 1}$, $E^{(2)}_{{\rm R}, 2}$ 
and $E^{(2)}_{{\rm R}, 3\pm}$, which are given by Eqs.~\eqref{eq:2-resonance-energy},
on the complex-$E$ plane in the case of $v^{\prime}=0$ and
(a) $\Delta U<4\Gamma$, (b) $\Delta U>4\Gamma$.}
\end{figure}

We next investigate the arrangement of the two eigenvalues $E^{(2)}_{{\rm R}, 3\pm}$ 
on the complex-$E$ plane in the simple case $v^{\prime}=0$. 
As is shown in the panel (a) of Fig.~\ref{fig:resonance-energies},
for $\Delta U<4\Gamma$, the two complex eigenvalues $E^{(2)}_{{\rm R}, 3\pm}$ 
share the same real part and have different imaginary parts 
that are symmetrically arranged with respect to the line $E=-2\ii\Gamma$ on the complex-$E$ plane.
At the exceptional point $\Delta U=4\Gamma$ giving $\xi=0$, 
the two eigenvalues $E^{(2)}_{{\rm R}, 3\pm}$ coalesce into one
and the matrix $S_{2}$ in Eq.~\eqref{eq:S-matrix_2} is not invertible since $p_{+}/q_{+}=p_{-}/q_{-}$,
whose discussion we defer to Section~\ref{sec:time-evolving-2-body-resonant-states_EP}.
For $\Delta U>4\Gamma$, the two eigenvalues $E^{(2)}_{{\rm R}, 3\pm}$ 
share the same imaginary part and have different real parts,
as is shown in the panel (b) of Fig.~\ref{fig:resonance-energies}.

We remark that, in the case of $U=U^{\prime}$ and $v^{\prime}=0$,
in which the two quantum dots are regarded as a single quantum dot with two degenerate energy levels,
the four eigenvalues become 
$E^{(2)}_{{\rm R},1}=E^{(2)}_{{\rm R},2}=2\epsilon_{\rm d}+U-2\ii\Gamma$,
$E^{(2)}_{{\rm R},3+}=2\epsilon_{\rm d}+U$ and
$E^{(2)}_{{\rm R},3-}=2\epsilon_{\rm d}+U-4\ii\Gamma$.
The two-body resonance energy $E^{(2)}_{{\rm R},3+}$ loses its imaginary part
and reduces to a bound-state energy of two electrons on the quantum dots.
In a way similar to the steady state with the real eigenvalue $E^{(1)}_{{\rm R}+}$ 
in Eq.~\eqref{eq:1-resonance-energy_2}, 
the emergence of the two-body bound state is understood by introducing the wave function
\begin{eqnarray}
 f_{{\rm BS},\sigma\overline{\sigma}}(t)
&=\frac{1}{2}\big(f_{11,\sigma\overline{\sigma}}(t)+f_{22,\sigma\overline{\sigma}}(t)
  -f_{12,\sigma\overline{\sigma}}(t)-f_{21,\sigma\overline{\sigma}}(t)\big)
\end{eqnarray}
that satisfies the decoupled differential equation
\begin{eqnarray}
&\big(\ii\partial_{t}-E^{(2)}_{{\rm R},3+}\big)f_{{\rm BS},\sigma\overline{\sigma}}(t)=0
\end{eqnarray}
without imposing the Siegert boundary conditions in Eq.~\eqref{eq:Siegert-2elec-bc_1}.

Let us finally solve the differential equation~\eqref{eq:Sch-eq_2elec-f_3}
for $\Delta U\neq 4\Gamma$.
By using the relation in Eq.~\eqref{eq:2-resonance-energy},
the differential equation~\eqref{eq:Sch-eq_2elec-f_3} is transformed to
\begin{eqnarray}
\label{eq:Sch-eq_2elec-f_4}
\fl
&S^{-1}_{2}S^{-1}_{1}
 \left(
 \begin{array}{c}
 \ii\partial_{t}f_{11,\sigma\overline{\sigma}}(t) \\
 \ii\partial_{t}f_{12,\sigma\overline{\sigma}}(t) \\
 \ii\partial_{t}f_{21,\sigma\overline{\sigma}}(t) \\
 \ii\partial_{t}f_{22,\sigma\overline{\sigma}}(t)
 \end{array}
 \right)
=\diag(E^{(2)}_{{\rm R}, 1}, E^{(2)}_{{\rm R}, 2}, E^{(2)}_{{\rm R}, 3+}, E^{(2)}_{{\rm R}, 3-})
 S^{-1}_{2}S^{-1}_{1}
 \left(
 \begin{array}{c}
 f_{11,\sigma\overline{\sigma}}(t) \\
 f_{12,\sigma\overline{\sigma}}(t) \\
 f_{21,\sigma\overline{\sigma}}(t) \\
 f_{22,\sigma\overline{\sigma}}(t)
 \end{array}
 \right)
\end{eqnarray}
which is readily solved as
\begin{eqnarray}
\label{eq:general-solution_f}
&\left(
 \begin{array}{c}
   f_{11,\sigma\overline{\sigma}}(t) \\
   f_{12,\sigma\overline{\sigma}}(t) \\
   f_{21,\sigma\overline{\sigma}}(t) \\
   f_{22,\sigma\overline{\sigma}}(t)
 \end{array}
 \right)
 =S_{1}S_{2}
  \left(
  \begin{array}{c}
   C_{1,\sigma\overline{\sigma}}\ee^{-\ii E^{(2)}_{{\rm R},1} t} \\
   C_{2,\sigma\overline{\sigma}}\ee^{-\ii E^{(2)}_{{\rm R},2} t} \\
   C_{3+,\sigma\overline{\sigma}}\ee^{-\ii E^{(2)}_{{\rm R},3+} t} \\
   C_{3-,\sigma\overline{\sigma}}\ee^{-\ii E^{(2)}_{{\rm R},3-} t}
  \end{array}
 \right).
\end{eqnarray}
Here $C_{1,\sigma\overline{\sigma}}$, $C_{2,\sigma\overline{\sigma}}$ and $C_{3\pm,\sigma\overline{\sigma}}$ 
are integration constants to be determined by the initial conditions of $f_{\alpha\beta,\sigma\overline{\sigma}}(t)$.
The wave function $f_{\alpha\beta,\sigma\overline{\sigma}}(t)$ at the exceptional point $\Delta U=4\Gamma$
shall be investigated in Section~\ref{sec:time-evolving-2-body-resonant-states_EP}.

\subsection{Time-evolving two-body resonant states}
\label{sec:time-evolving-2-body-resonant-states}

We now construct an exact time-evolving states by solving 
the set of time-dependent Schr\"odinger equations~\eqref{eq:Sch-eq_2elec-g_5},
\eqref{eq:Sch-eq_2elec-e_5} and \eqref{eq:Sch-eq_2elec-f_5}
for the initial state of localized two electrons on the two quantum dots,
which is given by
\begin{eqnarray}
\label{eq:initial-state}
 |\Psi_{\uparrow\downarrow}(0)\rangle
=\sum_{\alpha, \beta}\psi_{\alpha\beta,\uparrow\downarrow}
  d^{\dagger}_{\alpha\uparrow}d^{\dagger}_{\beta\downarrow}|0\rangle.
\end{eqnarray}
Here the coefficients $\psi_{\alpha\beta,\uparrow\downarrow}$ for $\alpha, \beta= 1, 2$
satisfy the Fermionic anti-symmetry relation 
$\psi_{\alpha\beta,\uparrow\downarrow}=-\psi_{\beta\alpha,\downarrow\uparrow}$ 
and the normalization condition 
$\sum_{\alpha, \beta}|\psi_{\alpha\beta,\uparrow\downarrow}|^{2}=1$.
In terms of wave functions, the initial state in Eq.~\eqref{eq:initial-state} is expressed as
\begin{eqnarray}
\label{eq:initial-condition}
 g_{m_{1}m_{2},\sigma\overline{\sigma}}(x_{1},x_{2},0)=0,\quad
 e_{m\alpha,\sigma\overline{\sigma}}(x,0)=0,\quad
 f_{\alpha\beta,\sigma\overline{\sigma}}(0)=\psi_{\alpha\beta,\sigma\overline{\sigma}}.
\end{eqnarray}
We solve this initial-value problem by the approach
that was developed in the previous work~\cite{Nishino-Hatano_24JPA}.

\begin{prop}
\label{prop:time-evolving-resonant-state}
{\rm
In the case $\Delta U\neq 4\Gamma$ in which there is no exceptional point,
the solution of the set of time-dependent Schr\"odinger equations~\eqref{eq:Sch-eq_2elec-g_5}, 
\eqref{eq:Sch-eq_2elec-e_5} and \eqref{eq:Sch-eq_2elec-f_5}
under the initial conditions in Eqs.~\eqref{eq:initial-condition} is given by
\numparts
\begin{eqnarray}
\label{eq:time-evolving-resonant-state_g}
\fl
&g_{m_{1}m_{2},\sigma\overline{\sigma}}(x_{1},x_{2},t)
 =-\frac{v^{2}}{2}\sum_{Q}\sum_{\alpha_{1},\alpha_{2}, s}
  \Big[1+\frac{(-1)^{\alpha_{1}+\alpha_{2}}\Delta U+4(v^{\prime}\!-\!\ii\Gamma)}{\xi_{s}}\Big]
  \psi_{\alpha_{Q_{1}}\alpha_{Q_{2}},\sigma\overline{\sigma}}
  \nn\\
\fl
&\hspace{97pt}\times
  \ee^{\ii E^{(2)}_{{\rm R},3s}(x_{Q_{2}}-t)-\ii E^{(1)}_{{\rm R}-}x_{Q_{2}Q_{1}}}
  \theta(t-x_{Q_{2}})\theta(x_{Q_{2}Q_{1}})\theta(x_{Q_{1}}),
  \\
\label{eq:time-evolving-resonant-state_e}
\fl
&e_{m\alpha,\sigma\overline{\sigma}}(x,t)
  =-\frac{\ii v}{2}\sum_{\beta}
  \big(\psi_{\beta\alpha,\sigma\overline{\sigma}}-\psi_{\overline{\beta}\overline{\alpha},\sigma\overline{\sigma}}\big)
  \ee^{\ii (\delta_{\beta\alpha}E^{(2)}_{{\rm R},1}+\delta_{\beta\overline{\alpha}}E^{(2)}_{{\rm R},2})(x-t)-\ii E^{(1)}_{{\rm R}+}x}\,
  \theta(t-x)\theta(x)
  \nn\\
\fl
&-\frac{\ii v}{4}\!\sum_{\alpha^{\prime}, \beta^{\prime}, s}\!
  \Big[1+\frac{(-1)^{\alpha^{\prime}+\beta^{\prime}}\Delta U+4(v^{\prime}\!-\!\ii\Gamma)}{\xi_{s}}\Big]
  \psi_{\alpha^{\prime}\beta^{\prime},\sigma\overline{\sigma}}\,
  \ee^{\ii E^{(2)}_{{\rm R},3s}(x-t)-\ii E^{(1)}_{{\rm R}-}x}\,\theta(t-x)\theta(x),
  \\
\label{eq:time-evolving-resonant-state_f}
\fl
&f_{\alpha\beta,\sigma\overline{\sigma}}(t)
  =\frac{1}{2}(\psi_{\alpha\beta,\sigma\overline{\sigma}}-\psi_{\overline{\alpha}\overline{\beta},\sigma\overline{\sigma}})
  \ee^{-\ii (\delta_{\alpha\beta}E^{(2)}_{{\rm R},1}+\delta_{\alpha\overline{\beta}}E^{(2)}_{{\rm R},2})t}
 \nn\\
\fl
&+\frac{1}{4}\sum_{s}\Big[\Big(1+\frac{(-1)^{\alpha+\beta}\Delta U}{\xi_{s}}\Big)
  (\psi_{\alpha\beta,\sigma\overline{\sigma}}+\psi_{\overline{\alpha}\overline{\beta},\sigma\overline{\sigma}})
  +\frac{4(v^{\prime}\!-\!\ii\Gamma)}{\xi_{s}}
  (\psi_{\alpha\overline{\beta},\sigma\overline{\sigma}}+\psi_{\overline{\alpha}\beta,\sigma\overline{\sigma}})\Big]
  \ee^{-\ii E^{(2)}_{{\rm R},3s}t}.
 \nn\\
\fl
&
\end{eqnarray}
\endnumparts
Here $Q=(Q_{1}, Q_{2})$ is a permutation of $(1, 2)$, $x_{12}=x_{1}-x_{2}$,
$E^{(1)}_{{\rm R}\pm}$ are the eigenvalues defined in Eq.~\eqref{eq:1-resonance-energy_2},
$E^{(2)}_{{\rm R},1}$, $E^{(2)}_{{\rm R}, 2}$ and $E^{(2)}_{{\rm R},3\pm}$
are  the two-body resonance energies defined in Eq.~\eqref{eq:2-resonance-energy}
and $\xi_{\pm}$ is defined in Eqs.~\eqref{eq:xi_def}.
}
\end{prop}

\begin{figure}
\begin{center}
{
\begin{picture}(360,135)(0,0)
\put(0,-15){\includegraphics[width=360pt]{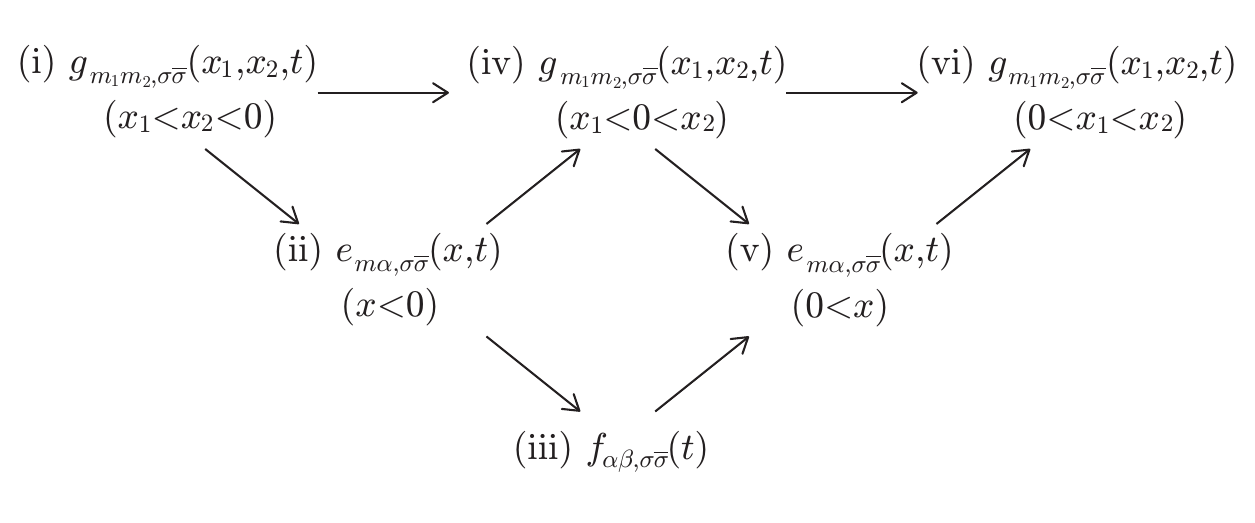}}
\end{picture}
}
\end{center}
\caption{\label{fig:flow-chart}
The flow chart of the construction of the time-evolving two-body resonant states
in Proposition~\ref{prop:time-evolving-resonant-state}.}
\end{figure}

\noindent
{\it Proof.}
We construct the time-evolving state following the flow chart in Fig.~\ref{fig:flow-chart}.

\noindent
(i):\; For $t>0$ and $x_{1}<x_{2}<0$, we have $x_{1}-t<x_{2}-t<0$. 
Then, by using the translation invariance in Eq.~\eqref{eq:translation-invariance_g}, 
we find 
\begin{equation}
 g_{m_{1}m_{2},\sigma\overline{\sigma}}(x_{1},x_{2},t)
 =g_{m_{1}m_{2},\sigma\overline{\sigma}}(x_{1}-t,x_{2}-t,0)=0.
\end{equation}

\noindent
(i)$\to$(ii):\; Because $g_{m_{1}m_{2},\sigma\overline{\sigma}}(x, 0-,t)=0$ for $x<0$,
the general solution of the wave function $e_{m\alpha,\sigma\overline{\sigma}}(x,t)$ is given by
Eq.~\eqref{eq:general-solution_e}.
Through the initial condition $e_{m\alpha,\sigma\overline{\sigma}}(x,0)=0$ 
in Eqs.~\eqref{eq:initial-condition}, we obtain $D_{m,\sigma\overline{\sigma},\pm}(x)=0$ for arbitrary $x<0$.
Since we have $x-t<0$ in the present case, we replace the variable $x$ of $D_{m,\sigma\overline{\sigma},\pm}(x)=0$
by $x-t$ and obtain $D_{m,\sigma\overline{\sigma},\pm}(x-t)=0$,
giving $e_{m\alpha,\sigma\overline{\sigma}}(x,t)=0$.

\noindent
(i), (ii)$\to$(iv):\;
For $x_{1}<0<x_{2}$, we have $x_{1}-x_{2}<0$.
Then, by using the translation invariance in Eq.~\eqref{eq:translation-invariance_g}
and the matching condition in Eq.~\eqref{eq:matching-cond_g}, 
we find
\begin{eqnarray}
\fl
&g_{m_{1}m_{2},\sigma\overline{\sigma}}(x_{1},x_{2},t)
 =g_{m_{1}m_{2},\sigma\overline{\sigma}}(x_{1}-x_{2},0+,t-x_{2})
  \nn\\
\fl
&=g_{m_{1}m_{2},\sigma\overline{\sigma}}(x_{1}-x_{2},0-,t-x_{2})
  -\ii v\sum_{\alpha}
  e_{m_{1}\alpha,\sigma\overline{\sigma}}(x_{1}-x_{2},t-x_{2})=0.
\end{eqnarray}

\noindent
(ii)$\to$(iii):\;
Because $e_{m\alpha,\sigma\overline{\sigma}}(x,t)=0$ for $x<0$,
the general solution of the wave function $f_{\alpha\beta,\sigma\overline{\sigma}}(t)$ is given by 
Eq.~\eqref{eq:general-solution_f}.
By imposing the initial conditions 
$f_{\alpha\beta,\sigma\overline{\sigma}}(0)=\psi_{\alpha\beta,\sigma\overline{\sigma}}$
in Eqs.~\eqref{eq:initial-condition} on the general solutions,
the integration constants in Eq.~\eqref{eq:general-solution_f} 
are determined as
\begin{eqnarray}
&\left(
  \begin{array}{c}
   C_{1,\sigma\overline{\sigma}} \\
   C_{2,\sigma\overline{\sigma}} \\
   C_{3+,\sigma\overline{\sigma}} \\
   C_{3-,\sigma\overline{\sigma}}
  \end{array} 
  \right)
  =S_{2}^{-1}S_{1}^{-1}
  \left(
  \begin{array}{c}
   \psi_{11,\sigma\overline{\sigma}} \\
   \psi_{12,\sigma\overline{\sigma}} \\
   \psi_{21,\sigma\overline{\sigma}} \\
   \psi_{22,\sigma\overline{\sigma}}
  \end{array}
  \right).
\end{eqnarray}
By inserting them into Eq.~\eqref{eq:general-solution_f}, 
the wave function $f_{\alpha\beta,\sigma\overline{\sigma}}(t)$ 
under the initial conditions in Eqs.~\eqref{eq:initial-condition} is obtained as
\begin{eqnarray}
\fl
&\left(
 \begin{array}{c}
  \! f_{11,\sigma\overline{\sigma}}(t)\! \\
  \! f_{12,\sigma\overline{\sigma}}(t)\! \\
  \! f_{21,\sigma\overline{\sigma}}(t)\! \\
  \! f_{22,\sigma\overline{\sigma}}(t)\!
 \end{array}
 \right)
 \!=\! S_{1}S_{2}\,
 \diag(\ee^{-\ii E^{(2)}_{{\rm R},1}t}, \ee^{-\ii E^{(2)}_{{\rm R},2}t}, \ee^{-\ii E^{(2)}_{{\rm R},3+}t}, \ee^{-\ii E^{(2)}_{{\rm R},3-}t})
 S_{2}^{-1}S_{1}^{-1}
 \left(
  \begin{array}{c}
  \!\psi_{11,\sigma\overline{\sigma}}\! \\
  \!\psi_{12,\sigma\overline{\sigma}}\! \\
  \!\psi_{21,\sigma\overline{\sigma}}\! \\
  \!\psi_{22,\sigma\overline{\sigma}}\!
 \end{array}
 \right).
\end{eqnarray}
Through the calculation of the products of the matrices with the relations
\begin{eqnarray}
&p_{+}p_{-}=-q_{+}q_{-}=-\frac{2(v^{\prime}-\ii\Gamma)}{\xi_{+}},\quad
  p_{\pm}q_{\mp}=\frac{-\Delta U+\xi_{\pm}}{2\xi_{+}},
\end{eqnarray}
the wave functions $f_{\alpha\beta,\sigma\overline{\sigma}}(t)$  for $\alpha, \beta=1, 2$
are summarized as Eq.~\eqref{eq:time-evolving-resonant-state_f}.

\noindent
(iv), (iii)$\to$(v):\;
We apply the general solution of the wave function 
$e_{m\alpha,\sigma\overline{\sigma}}(x,t)$ for $x>0$
in Eq.~\eqref{eq:general-solution_e} to the matching conditions at $x=0$ in Eq.~\eqref{eq:matching-cond_e}
with $v_{\rm F}=1$; we use the arbitrary functions $\tilde{D}_{m,\sigma\overline{\sigma}, \pm}(x-t)$
in place of $D_{m,\sigma\overline{\sigma}, \pm}(x-t)$ to distinguish from the case~(ii).
Then we have the following coupled equations for the functions $\tilde{D}_{m,\sigma\overline{\sigma}, \pm}(-t)$:
\begin{eqnarray}
\label{eq:coupled-eq_D_1}
\fl
&\tilde{D}_{m,\sigma\overline{\sigma},+}(-t)\ee^{-\ii E^{(1)}_{{\rm R}+}t/2}
 +\tilde{D}_{m,\sigma\overline{\sigma},-}(-t)\ee^{-\ii E^{(1)}_{{\rm R}-}t/2}
 =-\ii v\sum_{\beta}f_{\beta 1,\sigma\overline{\sigma}}(t),
 \nn\\
\fl
&-\tilde{D}_{m,\sigma\overline{\sigma},+}(-t)\ee^{-\ii E^{(1)}_{{\rm R}+}t/2}
 +\tilde{D}_{m,\sigma\overline{\sigma},-}(-t)\ee^{-\ii E^{(1)}_{{\rm R}-}t/2}
 =-\ii v\sum_{\beta}f_{\beta 2,\sigma\overline{\sigma}}(t).
\end{eqnarray}
By solving them and inserting the expression of $f_{\alpha\beta,\sigma\overline{\sigma}}(t)$ 
in Eq.~\eqref{eq:time-evolving-resonant-state_f} into the solution, we obtain
\begin{eqnarray}
\label{eq:coupled-eq_D_2}
\fl
\tilde{D}_{m,\sigma\overline{\sigma}, +}(-t)\ee^{-\ii E^{(1)}_{{\rm R}+}t/2}
&=\frac{v}{2\ii}\sum_{\beta}(\psi_{\beta 1,\sigma\overline{\sigma}}-\psi_{\overline{\beta}2,\sigma\overline{\sigma}})
  \ee^{-\ii (\delta_{\beta 1}E^{(2)}_{{\rm R},1}+\delta_{\beta 2}E^{(2)}_{{\rm R},2})t},
  \nn\\
\fl
\tilde{D}_{m,\sigma\overline{\sigma}, -}(-t)\ee^{-\ii E^{(1)}_{{\rm R}-}t/2}
&=\frac{v}{4\ii}\sum_{\alpha, \beta, s}\Big[\Big(1+\frac{(-1)^{\beta+\alpha}\Delta U}{\xi_{s}}\Big)
   \psi_{\beta\alpha,\sigma\overline{\sigma}}
  +\frac{4(v^{\prime}-\ii\Gamma)}{\xi_{s}}
   \psi_{\beta\overline{\alpha},\sigma\overline{\sigma}}\Big]
  \ee^{-\ii E^{(2)}_{{\rm R},3s}t}.
  \nn\\
\fl
&
\end{eqnarray}
We note that the relations in Eqs.~\eqref{eq:coupled-eq_D_2} hold for arbitrary $t>0$.
Since we have $t-x>0$ in the present case, we replace
the variable $t$ in Eqs.~\eqref{eq:coupled-eq_D_2} by $t-x$, 
resulting in
\begin{eqnarray}
\fl
&\tilde{D}_{m,\sigma\overline{\sigma}, +}(x-t)\ee^{-\ii E^{(1)}_{{\rm R}+}(t-x)/2}\theta(t-x)\theta(x)
 \nn\\
\fl
&=\frac{v}{2\ii}\sum_{\beta}
  \big(\psi_{\beta 1,\sigma\overline{\sigma}}-\psi_{\overline{\beta}2,\sigma\overline{\sigma}}\big)
  \ee^{-\ii (\delta_{\beta 1}E^{(2)}_{{\rm R},1}+\delta_{\beta 2}E^{(2)}_{{\rm R},2})(t-x)}\theta(t-x)\theta(x),
  \nn\\
\fl
&\tilde{D}_{m,\sigma\overline{\sigma}, -}(x-t)\ee^{-\ii E^{(1)}_{{\rm R}-}(t-x)/2}\theta(t-x)\theta(x)
  \nn\\
\fl
&=\frac{v}{4\ii}\sum_{\alpha, \beta, s}\Big[\Big(1+\frac{(-1)^{\beta+\alpha}\Delta U}{\xi_{s}}\Big)
   \psi_{\beta\alpha,\sigma\overline{\sigma}}
  +\frac{4(v^{\prime}-\ii\Gamma)}{\xi_{s}}
   \psi_{\beta\overline{\alpha},\sigma\overline{\sigma}}\Big]
  \ee^{-\ii E^{(2)}_{{\rm R},3s}(t-x)}\theta(t-x)\theta(x).
\end{eqnarray}
Here we have put the product of the step functions $\theta(t-x)\theta(x)$ 
on both sides of the equations in order to indicate that the relations hold only for $0<x<t$.
By inserting them into $e_{m\alpha,\sigma\overline{\sigma}}(x,t)$
in Eq.~\eqref{eq:general-solution_e}, we obtain Eq.~\eqref{eq:time-evolving-resonant-state_e}.

\noindent
(iv), (v)$\to$(vi):\;
Finally, we construct the wave function $g_{m_{1}m_{2},\sigma\overline{\sigma}}(x_{1},x_{2},t)$.
By applying the expression in Eq.~\eqref{eq:time-evolving-resonant-state_e}
to the first matching condition of $g_{m_{1}m_{2},\sigma\overline{\sigma}}(x_{1},x_{2},t)$ at $x_{1}=0$
in Eqs.~\eqref{eq:matching-cond_g} with $v_{\rm F}=1$ for $x_{2}>0$, we have
\begin{eqnarray}
\label{eq:matching-cond_g_2}
\fl
&g_{m_{1}m_{2},\sigma\overline{\sigma}}(0+,x_{2},t)
 =g_{m_{1}m_{2},\sigma\overline{\sigma}}(0-,x_{2},t)
  +\ii v\sum_{\alpha}e_{m_{2}\alpha,\overline{\sigma}\sigma}(x_{2},t)
 \nn\\
\fl
&=\frac{v^{2}}{2}\sum_{\alpha, \beta, s}
  \Big[\Big(1+\frac{(-1)^{\beta+\alpha}\Delta U}{\xi_{s}}\Big)
   \psi_{\beta\alpha,\overline{\sigma}\sigma}
  +\frac{4(v^{\prime}-\ii\Gamma)}{\xi_{s}}
   \psi_{\beta\overline{\alpha},\overline{\sigma}\sigma}\Big]
  \ee^{\ii E^{(2)}_{{\rm R},3s}(x_{2}-t)-\ii E^{(1)}_{{\rm R}-}x_{2}}\theta(t-x_{2}).
 \nn\\
\fl
&
\end{eqnarray}
By using the translation invariance in Eq.~\eqref{eq:translation-invariance_g} with $v_{\rm F}=1$
for $0<x_{1}<x_{2}$, we have
\begin{eqnarray}
&g_{m_{1}m_{2},\sigma\overline{\sigma}}(x_{1},x_{2},t)
 =g_{m_{1}m_{2},\sigma\overline{\sigma}}(0+,x_{2}-x_{1},t-x_{1}),
\end{eqnarray}
which gives Eq.~\eqref{eq:time-evolving-resonant-state_g} in the case $0<x_{1}<x_{2}$.
The wave function $g_{m_{1}m_{2},\sigma\overline{\sigma}}(x_{1},x_{2},t)$
in the case $0<x_{2}<x_{1}$ is obtained
through the anti-symmetry relations in Eqs.~\eqref{eq:anti-sym_g-f}.
\begin{flushright}
$\square$
\end{flushright}

We call the time-evolving state obtained in Proposition~\ref{prop:time-evolving-resonant-state}
{\it a time-evolving two-body resonant state}.
Similarly to the spinless case~\cite{Nishino-Hatano_24JPA},
the wave functions of the time-evolving two-body resonant state
in Eqs.~\eqref{eq:time-evolving-resonant-state_g}, \eqref{eq:time-evolving-resonant-state_e}
and \eqref{eq:time-evolving-resonant-state_f} decay exponentially in time
and grow exponentially in space, which is due to unbounded linear dispersion on the leads.
The former exponential decay in time of the wave function $f_{\alpha\beta,\sigma\overline{\sigma}}(t)$ on the quantum dots
is characterized by the two-body resonance energies
$E^{(2)}_{{\rm R},1}$, $E^{(2)}_{{\rm R},2}$ and $E^{(2)}_{{\rm R},3\pm}$ in Eq.~\eqref{eq:2-resonance-energy}.
The imaginary part of the two-body resonance energies 
determines the lifetime of the survival probability of two electrons on the two quantum dots,
which shall be described in Section~\ref{sec:survival-transition-prob}.

On the other hand, the latter exponential growth in space of the wave functions 
$g_{m_{1}m_{2},\sigma\overline{\sigma}}(x_{1},x_{2},t)$
is restricted to the interval $0<x_{1}, x_{2}<t$ and
that of $e_{m\alpha,\sigma\overline{\sigma}}(x,t)$ is also restricted to the interval $0<x<t$.
Both the space intervals expand in time with the electron velocity $v_{\rm F}=1$,
which is consistent with the causality.
As a result, the time-evolving resonant state in Proposition~\ref{prop:time-evolving-resonant-state} 
is normalizable at arbitrary time $t$ in contrast to the resonant states 
with spatially diverging wave functions of the time-independent Schr\"odinger equation~\cite{%
Hatano-Sasada-Nakamura-Petrosky_08PTP,Hatano-Kawamoto-Feinberg_09PJP}.

Here, due to the simplification of the system parameters,
the exponential decay in time of the wave functions $g_{m_{1}m_{2},\sigma\overline{\sigma}}(x_{1},x_{2},t)$
is characterized by the two-body resonance energies $E^{(2)}_{{\rm R},3\pm}$
with a $\Delta U$-dependent imaginary part,
and is not affected by $E^{(2)}_{{\rm R},1}$ and $E^{(2)}_{{\rm R},2}$.
Furtheremore, the wave functions $g_{m_{1}m_{2},\sigma\overline{\sigma}}(x_{1},x_{2},t)$ 
and $e_{m\alpha,\sigma\overline{\sigma}}(x,t)$ contain 
the exponential term $e^{-\ii E^{(1)}_{{\rm R} -}|x_{1}-x_{2}|}$ or $e^{-\ii E^{(1)}_{{\rm R} -}|x|}$
with the one-body resonance energy $E^{(1)}_{{\rm R} -}=\epsilon_{\rm d}+v^{\prime}-2\ii\Gamma$, 
which represents {\it two-body bound states} that decay exponentially 
with respect to the distance between the two electrons.
The binding strength of the two-body bound state is given by the imaginary part $2\Gamma$ 
of the one-body resonance energy $E^{(1)}_{{\rm R}-}$,
which was previously shown in the time-independent 
case~\cite{Nishino-Imamura-Hatano_12JPC,Nishino-Hatano-Ordonez_16JPC}.

\subsection{Two-body resonant states at an exceptional point}
\label{sec:time-evolving-2-body-resonant-states_EP}

We now investigate two-body resonant states at the exceptional point $\Delta U=4\Gamma$ with $v^{\prime}=0$ 
of the non-Hermite effective Hamiltonian $H^{(2)}_{\rm eff}$ in Eq.~\eqref{eq:eff-Hamiltonian_2-elec_2}.
As is indicated in Eqs.~\eqref{eq:2-resonance-energy},
at the exceptional point giving $\xi=0$, the two eigenvalues $E^{(2)}_{{\rm R}, 3\pm}$
coalesce into one and the corresponding eigenvectors become parallel.
In other words, the rank of the Hamiltonian matrix $H^{(2)}_{\rm eff}$ in Eq.~\eqref{eq:eff-Hamiltonian_2-elec_2}
decreases from four to three. Hence the Hamiltonian matrix $H^{(2)}_{\rm eff}$ 
in Eq.~\eqref{eq:eff-Hamiltonian_2-elec_2} is not diagonalizable with any similarity transformation.

Then we transform the Hamiltonian matrix $H^{(2)}_{\rm eff}$ into a Jordan normal form
in order to solve the differential equation~\eqref{eq:Sch-eq_2elec-f_3}.
Let us introduce
\begin{eqnarray}
&\tilde{S}_{2}=
 \left(
 \begin{array}{cccc}
   1 & 0 & 0 & 0 \\
   0 & 1 & 0 & 0 \\
   0 & 0 & 1 & -\frac{1}{\Delta U}  \\
   0 & 0 & \ii & \frac{\ii}{\Delta U} 
 \end{array}
 \right).
\end{eqnarray}
Through the similarity transformation of the effective Hamiltonian $H^{(2)}_{\rm eff}$ 
with the matrices $S_{1}$ in Eq.~\eqref{eq:orthogonal-matrix_1} and $\tilde{S}_{2}$, we have
\begin{eqnarray}
\fl
 \tilde{S}^{-1}_{2}S^{-1}_{1}H^{(2)}_{\rm eff}S_{1}\tilde{S}_{2}
 &=
  \left(
  \begin{array}{cccc}
   2\epsilon_{\rm d}+U-2\ii\Gamma & 0 & 0 & 0 \\
   0 & 2\epsilon_{\rm d}+U^{\prime}-2\ii\Gamma & 0 & 0 \\
   0 & 0 & 2\epsilon_{\rm d}+\overline{U}-\ii\frac{\Delta U}{2} &1  \\
   0 & 0 & 0 & 2\epsilon_{\rm d}+\overline{U}-\ii\frac{\Delta U}{2}
 \end{array}
 \right)
  \nn\\
\fl
&=:
 \left(
 \begin{array}{cccc}
   E^{(2)}_{{\rm R}, 1} & 0 & 0 & 0 \\
   0 & E^{(2)}_{{\rm R}, 2} & 0 & 0 \\
   0 & 0 & E^{(2)}_{{\rm R}, 3} &1  \\
   0 & 0 & 0 & E^{(2)}_{{\rm R}, 3}
 \end{array}
 \right),
\label{eq:Jordan-nomal-form}
\end{eqnarray}
which is in a Jordan normal form.
Here the eigenvalues $E^{(2)}_{{\rm R}, 1}$ and $E^{(2)}_{{\rm R}, 2}$ are
equal to the two-body resonance energies in Eq.~\eqref{eq:2-resonance-energy} 
for $\Delta U\neq 4\Gamma$
and the diagonal elements $E^{(2)}_{{\rm R}, 3}$ in the Jordan block agree with 
the resonance energy $E^{(2)}_{{\rm R}, 3\pm}$ in Eq.~\eqref{eq:2-resonance-energy} at $\xi=0$.

The differential equation~\eqref{eq:Sch-eq_2elec-f_3} is transformed to
\begin{eqnarray}
\fl
&\ii\partial_{t}\tilde{S}^{-1}_{2}S^{-1}_{1}
 \left(
 \begin{array}{c}
  f_{11,\sigma\overline{\sigma}}(t) \\
  f_{12,\sigma\overline{\sigma}}(t) \\
  f_{21,\sigma\overline{\sigma}}(t) \\
  f_{22,\sigma\overline{\sigma}}(t)
 \end{array}
 \right)
 =
 \left(
 \begin{array}{cccc}
   E^{(2)}_{{\rm R}, 1} & 0 & 0 & 0 \\
   0 & E^{(2)}_{{\rm R}, 2} & 0 & 0 \\
   0 & 0 & E^{(2)}_{{\rm R}, 3} &1  \\
   0 & 0 & 0 & E^{(2)}_{{\rm R}, 3}
 \end{array}
 \right)
 \tilde{S}^{-1}_{2}S^{-1}_{1}
 \left(
 \begin{array}{c}
  f_{11,\sigma\overline{\sigma}}(t) \\
  f_{12,\sigma\overline{\sigma}}(t) \\
  f_{21,\sigma\overline{\sigma}}(t) \\
  f_{22,\sigma\overline{\sigma}}(t)
 \end{array}
 \right).
\end{eqnarray}
Then, under the initial condition in Eq.~\eqref{eq:initial-condition},
it is solved as
\begin{eqnarray}
\fl
 \left(
 \begin{array}{c}
  \! f_{11,\sigma\overline{\sigma}}(t)\!  \\
  \! f_{12,\sigma\overline{\sigma}}(t)\!  \\
  \! f_{21,\sigma\overline{\sigma}}(t)\!  \\
  \! f_{22,\sigma\overline{\sigma}}(t)\! 
 \end{array}
 \right)
&\!\!=S_{1}\tilde{S}_{2}
 \left(
 \begin{array}{cccc}
   \!\!\ee^{-\ii E^{(2)}_{{\rm R}, 1}t}\!\! & 0 & 0 & 0 \\
   0 & \!\!\ee^{-\ii E^{(2)}_{{\rm R}, 2}t}\!\! & 0 & 0 \\
   0 & 0 & \!\!\ee^{-\ii E^{(2)}_{{\rm R}, 3}t}\!\! & -\ii t\ee^{-\ii E^{(2)}_{{\rm R}, 3}t}\!\!  \\
   0 & 0 & 0 & \!\!\ee^{-\ii E^{(2)}_{{\rm R}, 3}t}\!\!
 \end{array}
 \right)
 \!\tilde{S}^{-1}_{2}S^{-1}_{1}\!
 \left(
  \begin{array}{c}
   \psi_{11,\sigma\overline{\sigma}} \\
   \psi_{12,\sigma\overline{\sigma}} \\
   \psi_{21,\sigma\overline{\sigma}} \\
   \psi_{22,\sigma\overline{\sigma}}
  \end{array}
  \right).
\end{eqnarray}
Other wave functions $g_{m_{1}m_{2},\sigma\overline{\sigma}}(x_{1},x_{2},t)$ 
and $e_{m\alpha,\sigma\overline{\sigma}}(x,t)$ are constructed in a way similar to the previous subsections.
The final result is given by
\numparts
\label{eq:time-evolving-resonant-state_EP}
\begin{eqnarray}
\fl
&g_{m_{1}m_{2},\sigma\overline{\sigma}}(x_{1},x_{2},t)
 =-v^{2}\sum_{Q}\sum_{\alpha_{1}, \alpha_{2}}
  \big\{1-2\big[(-1)^{\alpha_{1}+\alpha_{2}}\ii+1\big]\Gamma t\big\}
  \psi_{\alpha_{Q_{1}}\alpha_{Q_{2}},\sigma\overline{\sigma}}
  \nn\\
\fl
&\times
  \ee^{\ii(2\epsilon_{\rm d}+\overline{U}-2\ii\Gamma)(x_{Q_{2}}-t)
  -\ii(\epsilon_{\rm d}-2\ii\Gamma)x_{Q_{2}Q_{1}}}
  \theta(t-x_{Q_{2}})\theta(x_{Q_{2}Q_{1}})\theta(x_{Q_{1}}),
  \\
\fl
&e_{m\alpha,\sigma\overline{\sigma}}(x,t)
  =\frac{v}{2\ii}\sum_{\beta}
  \big(\psi_{\beta\alpha,\sigma\overline{\sigma}}
  -\psi_{\overline{\beta}\overline{\alpha},\sigma\overline{\sigma}}\big)
  \ee^{\ii (2\epsilon_{\rm d}+U_{\alpha\beta}-2\ii\Gamma)(x-t)-\ii\epsilon_{\rm d}x}
  \theta(t-x)
  \nn\\
\fl
&+\frac{v}{2\ii}\sum_{\alpha^{\prime}, \beta^{\prime}}
  \big\{1-2\big[(-1)^{\alpha^{\prime}+\beta^{\prime}}\ii+1\big]\Gamma t\big\}
  \psi_{\alpha^{\prime}\beta^{\prime},\sigma\overline{\sigma}}
  \ee^{\ii(2\epsilon_{\rm d}+\overline{U}-2\ii\Gamma)(x-t)-\ii(\epsilon_{\rm d}-2\ii\Gamma)x}\theta(t-x),
  \\
\label{eq:time-evolving-resonant-state_EP_f}
\fl
&f_{\alpha\beta,\sigma\overline{\sigma}}(t)
 =\frac{1}{2}(\psi_{\alpha\beta,\sigma\overline{\sigma}}
  -\psi_{\overline{\alpha}\overline{\beta},\sigma\overline{\sigma}})
  \ee^{-\ii(2\epsilon_{\rm d}+U_{\alpha\beta}-2\ii\Gamma)t}
 \nn\\
\fl
&+\frac{1}{2}\big\{\big[1-(-1)^{\alpha+\beta}2\ii\Gamma t\big]
  (\psi_{\alpha\beta,\sigma\overline{\sigma}}+\psi_{\overline{\alpha}\overline{\beta},\sigma\overline{\sigma}})
  -2\Gamma t
  (\psi_{\alpha\overline{\beta},\sigma\overline{\sigma}}+\psi_{\overline{\alpha}\beta,\sigma\overline{\sigma}})\big\}
  \ee^{-\ii(2\epsilon_{\rm d}+\overline{U}-2\ii\Gamma)t},
\end{eqnarray}
\endnumparts
where $\overline{U}=(U+U^{\prime})/2$.
The three wave functions include exponential functions multiplied by a term linear in $t$,
which were also seen in the spinless case~\cite{Nishino-Hatano_24JPA}.
We notice that the wave functions are reproduced 
by taking the limit $\xi\to 0$ of Eqs.~\eqref{eq:time-evolving-resonant-state_g},
\eqref{eq:time-evolving-resonant-state_e} and \eqref{eq:time-evolving-resonant-state_f}.

\section{Survival and transition probabilities}
\label{sec:survival-transition-prob}

We now investigate the survival and transition probabilities of localized two electrons on the quantum dots
by using the time-evolving resonant state $|\Psi_{\uparrow\downarrow}(t)\rangle$
constructed in Proposition~\ref{prop:time-evolving-resonant-state}.
The survival probability of the initial state $|\Psi_{\uparrow\downarrow}(0)\rangle$ in Eq.~\eqref{eq:initial-state} 
is expressed as
\begin{eqnarray}
 Q(t)=|\langle\Psi_{\uparrow\downarrow}(0)|\Psi_{\uparrow\downarrow}(t)\rangle|^{2}
 =\Big|\sum_{\alpha, \beta}\psi^{\ast}_{\alpha\beta,\uparrow\downarrow}f_{\alpha\beta,\uparrow\downarrow}(t)\Big|^{2},
\end{eqnarray}
while the transition probability from the initial state $|\Psi_{\uparrow\downarrow}(0)\rangle$ to 
another two-electron state $|\Phi_{\uparrow\downarrow}\rangle=\sum_{\alpha, \beta}\phi_{\alpha\beta,\uparrow\downarrow}
d^{\dagger}_{\alpha\uparrow}d^{\dagger}_{\beta\downarrow}|0\rangle$ is expressed as
\begin{eqnarray}
 P(t)=|\langle\Phi_{\uparrow\downarrow}|\Psi_{\uparrow\downarrow}(t)\rangle|^{2}
 =\Big|\sum_{\alpha, \beta}\phi^{\ast}_{\alpha\beta,\uparrow\downarrow}f_{\alpha\beta,\uparrow\downarrow}(t)\Big|^{2}.
\end{eqnarray}
Hence their time dependence is determined by the exponential behavior
of the wave functions $f_{\alpha\beta,\uparrow\downarrow}(t)$ in Eq.~\eqref{eq:time-evolving-resonant-state_f}.
Which two-body resonance energies contribute to the exponential behavior
depends on the choice of the initial state $|\Psi_{\uparrow\downarrow}(0)\rangle$.
We consider four-types of the initial state
by setting the coefficients $\psi_{\alpha\beta,\uparrow\downarrow}$
of the initial state $|\Psi_{\uparrow\downarrow}(0)\rangle$ in Eq.~\eqref{eq:initial-state} as
\begin{eqnarray}
\label{eq:initial-states_so(4)}
\mbox{(I)}&\quad
 \psi_{11,\uparrow\downarrow}=-\psi_{22,\uparrow\downarrow}=\frac{1}{\sqrt{2}},\quad
 \psi_{12,\uparrow\downarrow}=\psi_{21,\uparrow\downarrow}=0,
 \nn\\
&\quad\Leftrightarrow\quad
 |\Psi^{\rm (I)}_{\uparrow\downarrow}(0)\rangle=\frac{1}{\sqrt{2}}\big( 
 d^{\dagger}_{1\uparrow}d^{\dagger}_{1\downarrow}-d^{\dagger}_{2\uparrow}d^{\dagger}_{2\downarrow}\big)|0\rangle,
 \nn\\
\mbox{(II)}&\quad 
 \psi_{11,\uparrow\downarrow}=\psi_{22,\uparrow\downarrow}=0,\quad
 \psi_{12,\uparrow\downarrow}=-\psi_{21,\uparrow\downarrow}=\frac{1}{\sqrt{2}}
 \nn\\
&\quad\Leftrightarrow\quad
 |\Psi^{\rm (II)}_{\uparrow\downarrow}(0)\rangle=\frac{1}{\sqrt{2}}\big(
 d^{\dagger}_{1\uparrow}d^{\dagger}_{2\downarrow}-d^{\dagger}_{2\uparrow}d^{\dagger}_{1\downarrow}\big)|0\rangle,
 \nn\\
\mbox{(III)}&\quad 
 \psi_{11,\uparrow\downarrow}=\psi_{22,\uparrow\downarrow}=0,\quad
 \psi_{12,\uparrow\downarrow}=\psi_{21,\uparrow\downarrow}=\frac{1}{\sqrt{2}}
 \nn\\
&\quad\Leftrightarrow\quad
 |\Psi^{\rm (III)}_{\uparrow\downarrow}(0)\rangle=\frac{1}{\sqrt{2}}\big(
 d^{\dagger}_{1\uparrow}d^{\dagger}_{2\downarrow}+d^{\dagger}_{2\uparrow}d^{\dagger}_{1\downarrow}\big)|0\rangle,
 \nn\\
\mbox{(IV)}&\quad 
 \psi_{11,\uparrow\downarrow}=\psi_{22,\uparrow\downarrow}=\frac{1}{\sqrt{2}},\quad
 \psi_{12,\uparrow\downarrow}=\psi_{21,\uparrow\downarrow}=0
 \nn\\
&\quad\Leftrightarrow\quad
 |\Psi^{\rm (IV)}_{\uparrow\downarrow}(0)\rangle=\frac{1}{\sqrt{2}}\big(
 d^{\dagger}_{1\uparrow}d^{\dagger}_{1\downarrow}+d^{\dagger}_{2\uparrow}d^{\dagger}_{2\downarrow}\big)|0\rangle.
\end{eqnarray}
Here we let $|\Psi^{(\nu)}_{\uparrow\downarrow}(0)\rangle$ denote
the initial state $|\Psi_{\uparrow\downarrow}(0)\rangle$ in the case $(\nu)$ 
for $\nu=\mbox{I}, \mbox{II}, \mbox{III}, \mbox{IV}$.
The initial states $|\Psi^{\rm (I)}_{\uparrow\downarrow}(0)\rangle$
and $|\Psi^{\rm (IV)}_{\uparrow\downarrow}(0)\rangle$ are 
two-electron states of double occupancy on one of the two quantum dots,
while $|\Psi^{\rm (II)}_{\uparrow\downarrow}(0)\rangle$
and $|\Psi^{\rm (III)}_{\uparrow\downarrow}(0)\rangle$ are 
those of simultaneous occupancy on the two quantum dots.
The two states $|\Psi^{\rm (I)}_{\uparrow\downarrow}(0)\rangle$ 
and $|\Psi^{\rm (IV)}_{\uparrow\downarrow}(0)\rangle$
as well as the two states $|\Psi^{\rm (II)}_{\uparrow\downarrow}(0)\rangle$ 
and $|\Psi^{\rm (III)}_{\uparrow\downarrow}(0)\rangle$
are distinguished by the symmetry under the exchange of the two quantum dots.
We note that each initial state in Eqs.~\eqref{eq:initial-states_so(4)} corresponds to 
an irreducible representation of the semisimple Lie algebra $so(4)$,
which shall be described in \ref{sec:algebraic-structure}.

By inserting the coefficients $\psi_{\alpha\beta,\uparrow\downarrow}$ 
in each case of Eqs.~\eqref{eq:initial-states_so(4)}
into the wave functions in Eq.~\eqref{eq:time-evolving-resonant-state_f} for $\xi\neq 0$, 
we have 
\begin{eqnarray}
\label{eq:time-evolving-resonant-state_so(4)}
\fl
\mbox{(I)}\quad
&f^{\rm (I)}_{\alpha\alpha,\uparrow\downarrow}(t)
 =(-1)^{\overline{\alpha}}\frac{1}{\sqrt{2}}\ee^{-\ii E^{(2)}_{{\rm R},1}t},\quad
 f^{\rm (I)}_{\alpha\overline{\alpha},\uparrow\downarrow}(t)=0,
 \nn\\
\fl
\mbox{(II)}\quad
&f^{\rm (II)}_{\alpha\alpha,\uparrow\downarrow}(t)=0,\quad
 f^{\rm (II)}_{\alpha\overline{\alpha},\uparrow\downarrow}(t)
 =(-1)^{\overline{\alpha}}\frac{1}{\sqrt{2}}\ee^{-\ii E^{(2)}_{{\rm R},2}t},
 \nn\\
\fl
\mbox{(III)}\quad
&f^{\rm (III)}_{\alpha\alpha,\uparrow\downarrow}(t)
 =\sum_{s}\frac{\sqrt{2}(v^{\prime}-\ii\Gamma)}{\xi_{s}}
   \ee^{-\ii E^{(2)}_{{\rm R},3s}t},\quad
 f^{\rm (III)}_{\alpha\overline{\alpha},\uparrow\downarrow}(t)
 =\sum_{s}\frac{1}{2\sqrt{2}}\Big(1-\frac{\Delta U}{\xi_{s}}\Big)
  \ee^{-\ii E^{(2)}_{{\rm R},3s}t},
 \nn\\
\fl
\mbox{(IV)}\quad
&f^{\rm (IV)}_{\alpha\alpha,\uparrow\downarrow}(t)
 =\sum_{s}\frac{1}{2\sqrt{2}}\Big(1+\frac{\Delta U}{\xi_{s}}\Big)
   \ee^{-\ii E^{(2)}_{{\rm R},3s}t},
 \quad
 f^{\rm (IV)}_{\alpha\overline{\alpha},\uparrow\downarrow}(t)
 =\sum_{s}\frac{\sqrt{2}(v^{\prime}-\ii\Gamma)}{\xi_{s}}
   \ee^{-\ii E^{(2)}_{{\rm R},3s}t}.
 \nn\\
\fl
&
\end{eqnarray}
Here we let $f^{(\nu)}_{\alpha\beta,\uparrow\downarrow}(t)$ denote
the wave function $f_{\alpha\beta,\uparrow\downarrow}(t)$ 
with the coefficients $\psi_{\alpha\beta,\uparrow\downarrow}$ in the case $(\nu)$
of Eqs.~\eqref{eq:initial-states_so(4)} for $\nu=\mbox{I}, \mbox{II}, \mbox{III}, \mbox{IV}$.
We find that the exponential decay of the wave function $f^{\rm (I)}_{\alpha\beta,\uparrow\downarrow}(t)$ 
is determined only by the resonance energy $E^{(2)}_{{\rm R},1}$,
and that of $f^{\rm (II)}_{\alpha\beta,\uparrow\downarrow}(t)$ 
is determined only by $E^{(2)}_{{\rm R},2}$.
The inverse lifetime in both the cases is $2\Gamma$ since the resonance energies $E^{(2)}_{{\rm R},1}$
and $E^{(2)}_{{\rm R},2}$ share the same imaginary part.
We notice that the wave function $f^{\rm (II)}_{\alpha\beta,\uparrow\downarrow}(t)$ 
is the same as that in the spinless case~\cite{Nishino-Hatano_24JPA},
which is understood by the $so(4)$-algebraic structure of the effective Hamiltonian $H_{\rm eff}$
(see \ref{sec:algebraic-structure}).
On the other hand, the exponential decay of both the wave functions 
$f^{\rm (III)}_{\alpha\beta,\uparrow\downarrow}(t)$ 
and $f^{\rm (IV)}_{\alpha\beta,\uparrow\downarrow}(t)$
is determined by the two resonance energies $E^{(2)}_{{\rm R},3\pm}$
with the imaginary part $2\Gamma\pm\mathrm{Im}(\xi)/2$.
Hence, for $\mathrm{Im}(\xi)\neq 0$,
the wave functions $f^{\rm (III)}_{\alpha\beta,\uparrow\downarrow}(t)$ 
and $f^{\rm (IV)}_{\alpha\beta,\uparrow\downarrow}(t)$ exhibit 
the interference of two types of the exponential decay,
whereas those for $\mathrm{Im}(\xi)=0$ with $\xi\neq 0$ behave as a simple exponential decay
as is similar to $f^{\rm (I)}_{\alpha\beta,\uparrow\downarrow}(t)$ 
and $f^{\rm (II)}_{\alpha\beta,\uparrow\downarrow}(t)$.

At the exceptional point $\Delta U=4\Gamma$ with $v^{\prime}=0$,
the wave functions $f^{(\nu)}_{\alpha\beta,\uparrow\downarrow}(t)$
in the cases $\nu=\mbox{III}, \mbox{IV}$ are given by
\begin{eqnarray}
\fl
\mbox{(III)}\quad
&f^{\rm (III)}_{\alpha\alpha,\uparrow\downarrow}(t)=
 -\sqrt{2}\Gamma t
 \ee^{-\ii(2\epsilon_{\rm d}+\overline{U}-2\ii\Gamma)t},\quad
 f^{\rm (III)}_{\alpha\overline{\alpha},\uparrow\downarrow}(t)=
 \frac{1}{\sqrt{2}}(1+2\ii\Gamma t)
 \ee^{-\ii(2\epsilon_{\rm d}+\overline{U}-2\ii\Gamma)t},
 \nn\\
\fl
\mbox{(IV)}\quad
&f^{\rm (IV)}_{\alpha\alpha,\uparrow\downarrow}(t)=
 \frac{1}{\sqrt{2}}(1-2\ii\Gamma t)
  \ee^{-\ii(2\epsilon_{\rm d}+\overline{U}-2\ii\Gamma)t},\quad
 f^{\rm (IV)}_{\alpha\overline{\alpha},\uparrow\downarrow}(t)=
 -\sqrt{2}\Gamma t
  \ee^{-\ii(2\epsilon_{\rm d}+\overline{U}-2\ii\Gamma)t}.
\end{eqnarray}
We find that the diagonal element $E^{(2)}_{{\rm R}, 3}$
of the Jordan block in Eq.~\eqref{eq:Jordan-nomal-form} appears
in the exponential functions multiplied by a term linear in $t$.

Now, we explicitly calculate the survival and transition probabilities
of the initial states $|\Psi^{(\nu)}_{\uparrow\downarrow}(0)\rangle$ 
for $\nu=\mbox{I}, \mbox{II}, \mbox{III}, \mbox{IV}$ in Eqs.~\eqref{eq:initial-states_so(4)}.
Let $Q^{(\nu)}(t)$ denote the survival probability of the initial state $|\Psi^{(\nu)}_{\uparrow\downarrow}(0)\rangle$,
and $P^{(\nu)\to(\mu)}(t)$ the transition probability from the initial state $|\Psi^{(\nu)}_{\uparrow\downarrow}(0)\rangle$
to the final state $|\Psi^{(\mu)}_{\uparrow\downarrow}(0)\rangle$ for $\mu\neq\nu$.
In what follows, we consider the case $v^{\prime}=0$
in order to investigate the change of time evolution around the exceptional point 
at $\Delta U=4\Gamma$.

The survival probabilities $Q^{(\nu)}(t)$ in the case $v^{\prime}=0$ are calculated as
\begin{eqnarray}
\label{eq:survival-prob_12}
\fl
&Q^{\rm (I)}(t)=Q^{\rm (II)}(t)=\ee^{-4\Gamma t},
 \\
\label{eq:survival-prob_34}
\fl
&Q^{\rm (III)}(t)=Q^{\rm (IV)}(t)=
 \cases{
  \Big[1+\frac{16\Gamma^{2}}{\eta^{2}}\sinh^{2}\Big(\frac{\eta}{2}t\Big)\Big]\ee^{-4\Gamma t}
 & for $\Delta U<4\Gamma$, \\
  (1+4\Gamma^{2}t^{2})\ee^{-4\Gamma t}
 & for $\Delta U=4\Gamma$, \\
  \Big[1+\frac{16\Gamma^{2}}{\xi^{2}}\sin^{2}\Big(\frac{\xi}{2}t\Big)\Big]\ee^{-4\Gamma t}
 & for $\Delta U>4\Gamma$,
 }
\end{eqnarray}
where $\xi$ in Eq.~\eqref{eq:xi_def} becomes $\xi=\sqrt{(\Delta U)^{2}-16\Gamma^{2}}$ 
in the case $v^{\prime}=0$ and
we introduce $\eta:=\mathrm{Im}(\xi)=\sqrt{16\Gamma^{2}-(\Delta U)^{2}}$ for $\Delta U<4\Gamma$.
Both the survival probabilities $Q^{\rm (I)}(t)$ and $Q^{\rm (II)}(t)$
in Eq.~\eqref{eq:survival-prob_12} 
are independent of the interactions $U$ and $U^{\prime}$
and decay exponentially in time with the same inverse lifetime $4\Gamma$,
which is described by the contribution of the resonance energies $E^{(2)}_{{\rm R}, 1}$ and $E^{(2)}_{{\rm R}, 2}$
in Eq.~\eqref{eq:2-resonance-energy}, respectively.
As was already mentioned for the wave functions in Eq.~\eqref{eq:time-evolving-resonant-state_so(4)},
the inverse lifetime $4\Gamma$ is the same as that 
in the case of two spinless electrons~\cite{Nishino-Hatano_24JPA}.
On the other hand, the survival probabilities $Q^{\rm (III)}(t)$ and $Q^{\rm (IV)}(t)$ in Eq.~\eqref{eq:survival-prob_34}
are affected by the interference of the two resonance energies $E^{(2)}_{{\rm R}, 3\pm}$ 
in Eq.~\eqref{eq:2-resonance-energy}, and are classified into three cases depending on $\Delta U$.
For $\Delta U<4\Gamma$, the survival probabilities $Q^{\rm (III)}(t)$ and $Q^{\rm (IV)}(t)$ 
decay exponentially in time with the inverse lifetime $4\Gamma-\eta$ 
while for $\Delta U>4\Gamma$, it oscillates during the exponential decay with the inverse lifetime $4\Gamma$.
At the exceptional point at $\Delta U=4\Gamma$, 
the survival probabilities $Q^{\rm (III)}(t)$ and $Q^{\rm (IV)}(t)$
decay in the form of an exponential function multiplied by a quadratic function in $t$.

\begin{figure}[h]
\begin{center}
{
\begin{picture}(320,200)(0,0)
\put(0,-5){\includegraphics[width=320pt]{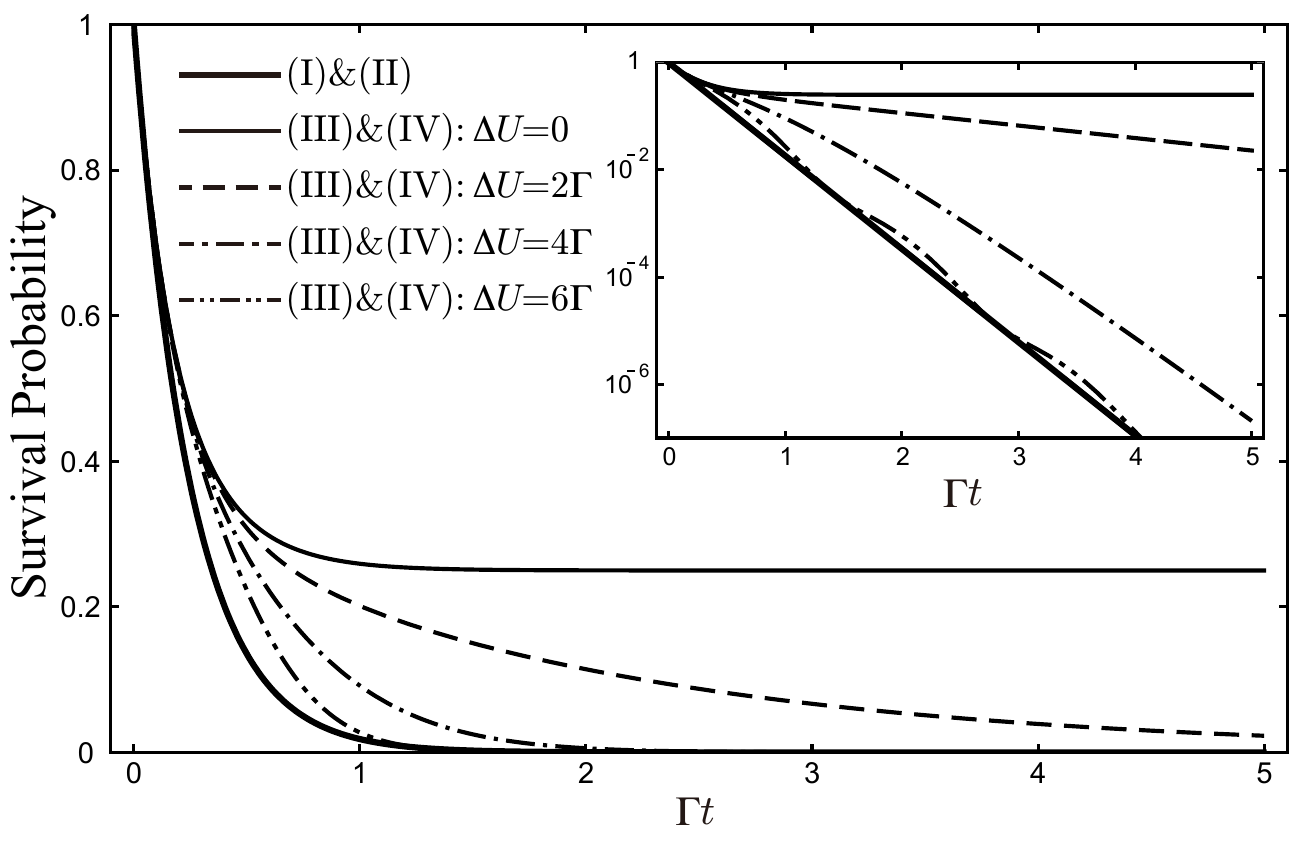}}
\end{picture}
}
\end{center}
\caption{\label{fig:SP_1234} 
Time-dependence of the survival probabilities 
$Q^{\rm (I)}(t)$ and $Q^{\rm (II)}(t)$ for arbitrary $U$ and $U^{\prime}$,
and $Q^{\rm (III)}(t)$ and $Q^{\rm (IV)}(t)$ for $\Delta U=0, 2\Gamma, 4\Gamma, 6\Gamma$.
The case $\Delta U=4\Gamma$ of $Q^{\rm (III)}(t)$ and $Q^{\rm (IV)}(t)$
corresponds to the exceptional point of the effective Hamiltonian $H^{(2)}_{\rm eff}$
in Eq.~\eqref{eq:eff-Hamiltonian_2-elec_2}.
Their semi-logarithmic plots are presented in the inset to show
the oscillation in time in the case of $\Delta U=6\Gamma$ of $Q^{\rm (III)}(t)$ and $Q^{\rm (IV)}(t)$.}
\end{figure}

Figure~\ref{fig:SP_1234} shows the time-dependence of the survival probabilities 
$Q^{\rm (I)}(t)$ and $Q^{\rm (II)}(t)$ for arbitrary $U$ and $U^{\prime}$, 
and $Q^{\rm (III)}(t)$ and $Q^{\rm (IV)}(t)$ for $\Delta U=0, 2\Gamma, 4\Gamma, 6\Gamma$.
Recall that the survival probabilities $Q^{\rm (III)}(t)$ and $Q^{\rm (IV)}(t)$ in Eqs.~\eqref{eq:survival-prob_34}
depend not on the interaction $U$ nor $U^{\prime}$ directly but on the difference $\Delta U$.
The survival probabilities $Q^{\rm (I)}(t)$ and $Q^{\rm (II)}(t)$
show a purely exponential decay in time.
In the case $\Delta U=0$, the survival probabilities $Q^{\rm (III)}(t)$ and $Q^{\rm (IV)}(t)$
converge to the value $1/4$ in the long-time limit $t\to\infty$,
which is due to the emergence of a bound state
with the real energy eigenvalue $E^{(2)}_{{\rm R}, 3+}=2\epsilon_{\rm d}+U$.
For $\Delta U>0$, they decay exponentially in time,
whose lifetime is longer than that of $Q^{\rm (I)}(t)$ and $Q^{\rm (II)}(t)$.
In order to show the oscillation of $Q^{\rm (III)}(t)$ and $Q^{\rm (IV)}(t)$ in the case $\Delta U=6\Gamma$,
we present a semi-logarithmic plot of them in the inset of Fig.~\ref{fig:SP_1234}.

The transition probabilities $P^{(\nu)\to(\mu)}(t)$ in the case $v^{\prime}=0$ are calculated as
\begin{eqnarray}
\label{eq:transition-prob_12}
\fl
&P^{(\nu)\to(\mu)}(t)=0\quad 
 \mbox{for } (\nu, \mu)\neq (\mbox{III}, \mbox{IV}), (\mbox{IV}, \mbox{III})
 \\
\label{eq:transition-prob_34}
\fl
&P^{{\rm (III)}\to{\rm (IV)}}(t)=P^{{\rm (IV)}\to{\rm (III)}}(t)
 =
 \cases{
  \frac{16\Gamma^{2}}{\eta^{2}}\sinh^{2}\Big(\frac{\eta}{2}t\Big)
  \ee^{-4\Gamma t}
 & for $\Delta U<4\Gamma$, \\
  4\Gamma^{2}t^{2}
  \ee^{-4\Gamma t}
 & for $\Delta U=4\Gamma$, \\
  \frac{16\Gamma^{2}}{\xi^{2}}\sin^{2}\Big(\frac{\xi}{2}t\Big)
  \ee^{-4\Gamma t}
 & for  $\Delta U>4\Gamma$.
 }
\end{eqnarray}
The results in Eq.~\eqref{eq:transition-prob_12}
indicate that the initial states $|\Psi^{\rm (I)}_{\uparrow\downarrow}(0)\rangle$
and $|\Psi^{\rm (II)}_{\uparrow\downarrow}(0)\rangle$ decay directly to the external leads  
without being transferred to other states on the quantum dots 
while Eq.~\eqref{eq:transition-prob_34} indicates that
the two initial states $|\Psi^{\rm (III)}_{\uparrow\downarrow}(0)\rangle$ 
and $|\Psi^{\rm (IV)}_{\uparrow\downarrow}(0)\rangle$ are transferred to
each other during the decay to the leads.
The transfer between the two states $|\Psi^{\rm (III)}_{\uparrow\downarrow}(0)\rangle$ 
and $|\Psi^{\rm (IV)}_{\uparrow\downarrow}(0)\rangle$ is consistent with
the block-diagonal structure of the effective Hamiltonian $H^{(2)}_{\rm eff}$
in Eq.~\eqref{eq:eff-Hamiltonian_2-elec_3}.
As is similar to the survival probabilities,
the time-dependence of the transition probabilities $P^{{\rm (III)}\to{\rm (IV)}}(t)$ and $P^{{\rm (IV)}\to{\rm (III)}}(t)$
is classified into three cases depending on $\Delta U$.

\begin{figure}[h]
\begin{center}
{
\begin{picture}(320,200)(0,0)
\put(0,-5){\includegraphics[width=320pt]{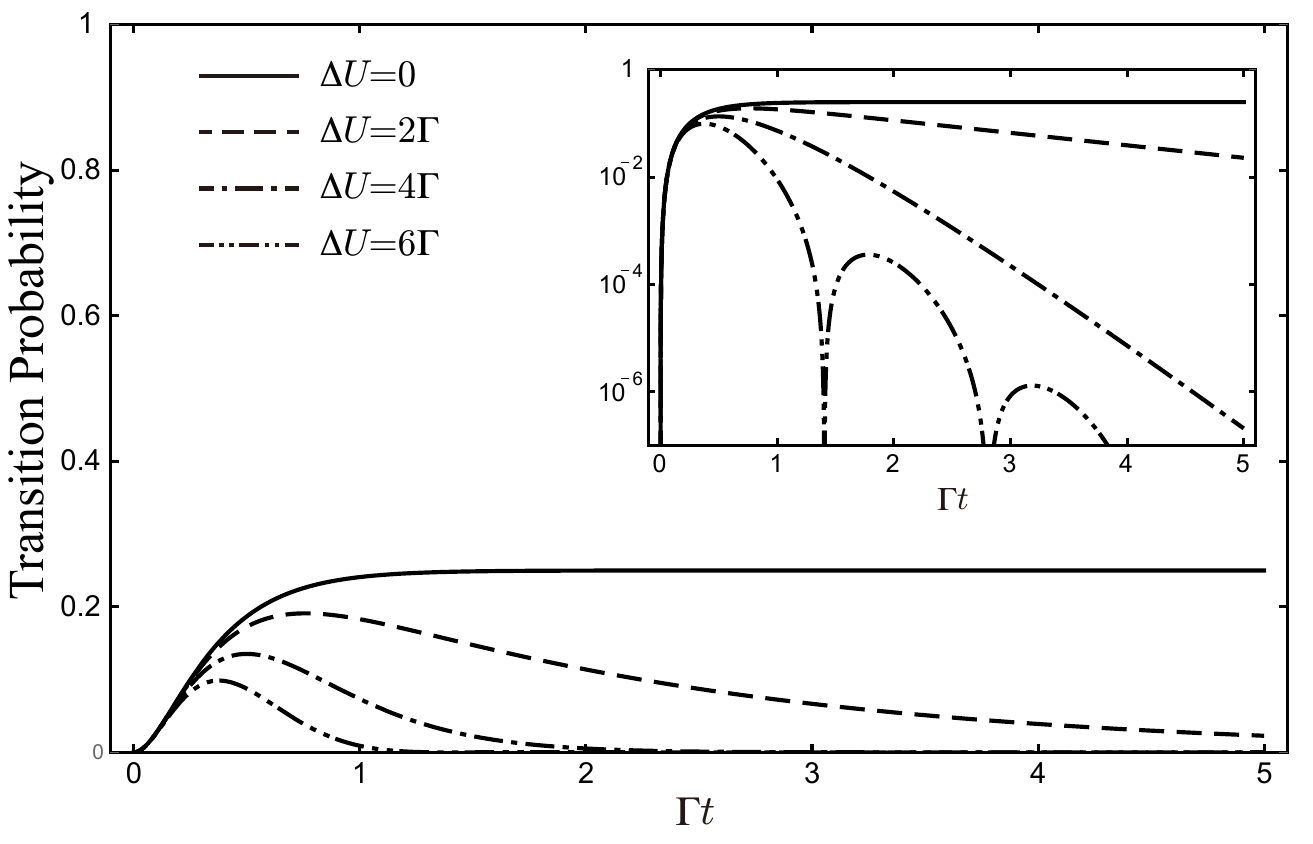}}
\end{picture}
}
\end{center}
\caption{\label{fig:TP_34} 
Time-dependence of the transition probabilities $P^{{\rm (III)}\to{\rm (IV)}}(t)$
from the initial state $|\Psi^{\rm (III)}_{\uparrow\downarrow}(0)\rangle$ 
to a final state $|\Psi^{\rm (IV)}_{\uparrow\downarrow}(0)\rangle$ for $\Delta U=0, 2\Gamma, 4\Gamma, 6\Gamma$.
The case $\Delta U=4\Gamma$ corresponds to the exceptional point.
Their semi-logarithmic plots are presented in the inset to show
the oscillation in time in the case of $\Delta U=6\Gamma$.}
\end{figure}

Figure~\ref{fig:TP_34} shows the time-dependence of the transition probabilities $P^{{\rm (III)}\to{\rm (IV)}}(t)$
and $P^{{\rm (IV)}\to{\rm (III)}}(t)$ for $\Delta U=0, 2\Gamma, 4\Gamma, 6\Gamma$.
The initial increase of the transition probabilities $P^{{\rm (III)}\to{\rm (IV)}}(t)$ and $P^{{\rm (IV)}\to{\rm (III)}}(t)$
indicates the transition between the two initial states $|\Psi^{\rm (III)}_{\uparrow\downarrow}(0)\rangle$ 
and $|\Psi^{\rm (IV)}_{\uparrow\downarrow}(0)\rangle$ for all cases.
Only in the case $\Delta U=0$, they converge to the value $1/4$ in the long-time limit $t\to\infty$.
This result together with the convergence of the survival probability to $1/4$ implies 
that the total probability for the two initial states to survive on the quantum dots converges to $1/2$ in the limit  $t\to\infty$.
For $\Delta U>0$, the transition probabilities decay exponentially after the initial increase.
We present a semi-logarithmic plot of the transition probabilities in the inset of Fig.~\ref{fig:TP_34}
in order to show the oscillation in time for $\Delta U=6\Gamma$.

\section{Concluding remarks}
\label{sec:concluding-remarks}

We have studied time-evolving resonant states in an open double quantum-dot system 
with spin degrees of freedom as well as both on-dot and interdot Coulomb interactions.
By extending the Siegert boundary conditions to the two-electron case,
we have derived non-Hermite effective Hamiltonians exactly,
whereby we have obtained two-body resonance energies that depend on interaction parameters.

We have constructed exact time-evolving two-body resonant states by solving the time-dependent
Schr\"odinger equation under the initial condition of localized two electrons on the quantum dots.
The exact solution indicates that the wave functions on the quantum dots decay exponentially in time
and those on the external leads grow exponentially 
within a finite space interval that expands in time with the electron velocity.
The exact solution also enables the calculation of the survival and transition probabilities
of localized two electrons on the quantum dots.
The decay in time of the probabilities is classified by taking initial states 
based on the $so(4)$-algebraic structure of the non-Hermite effective Hamiltonian.
We note that our construction is restricted to systems with linear dispersion relations,
which allows for an exact treatment of the interactions.
In contrast, the Green's function method~\cite{Hatano-Ordonez_14JMP,Ordonez-Hatano_17JPA,%
Ordonez-Hatano_17Chaos,Hatano_21JPCS,Feshbach_58ARNS,Feshbach_58AP,Feshbach_62AP,%
Garmon-Petrosky-Simine-Segal_13FortschrPhys,%
Chakraborty-Sensarma_18PRB,Garmon-Noba-Ordonez-Segal_19PRD}
is applicable to systems with non-linear dispersion relations, but in interacting cases, 
it typically requires perturbative calculations.
We expect that the results that we exactly obtained for linear dispersion relations 
will serve as reference points for calculations by the Green's function method for non-linear dispersion relations.

The purely exponential behavior of the time-evolving resonant states 
and the survival and transition probabilities is due to unbounded linear dispersion on the external leads. 
It was shown in the study of a quantum Zeno effect
that the lower limit of the dispersion would result in deviations from 
the exponential behavior in a short-time regime~\cite{%
Khalfin_68PZETF,Chiu-Sudarshan-Misra_77PRD,Petrosky-Tasaki-Prigogine_91PhysicaA,Petrosky-Ordonez-Prigogine_01PRA,%
Garmon-Petrosky-Simine-Segal_13FortschrPhys,Chakraborty-Sensarma_18PRB,Garmon-Noba-Ordonez-Segal_19PRD}.
Quite recently, a collaborator and 
we have found that the Dirac mass-gap in unbounded dispersion relations induces a power-law decay of 
the survival probability in the long-time regime while preserving the exponential behavior 
in the short-time regime~\cite{Taira-Hatano-Nishino_2024preprint}.
The strict step-function behavior in which the wave functions grow exponentially only within a finite interval
is also due to the unbounded linear dispersion.
Such behavior of the wave functions should be reproduced in the infinite-band limit of 
the time-evolving wave functions of open quantum systems 
with a finite-band dispersion~\cite{Petrosky-Ordonez-Prigogine_01PRA,Hatano-Ordonez_14JMP}.

It is interesting to compare the non-Hermite effective Hamiltonian
obtained by imposing the Siegert boundary conditions on the Schr\"odinger equation 
with those derived from the Feshbach formalism~\cite{%
Feshbach_58ARNS,Feshbach_58AP,Feshbach_62AP}.
For some open quantum-dot systems without interactions, 
the effective Hamiltonians obtained by the two approaches
are shown to be identical~\cite{Hatano-Ordonez_14JMP,Hatano_21JPCS,Sasada-Hatano_08JPSJ}.
The equivalence of the two approaches indicates that the imaginary part of the effective Hamiltonian 
corresponds to the self-energy including effects of the leads, 
which is independent of energies for the present systems with linear dispersion relations.

\section*{Acknowledgments}

The authors thank Prof.~T.~Y.~Petrosky, Prof.~G.~Ordonez and Prof.~S.~Garmon for helpful comments.
N.H.'s study is partially supported by Japan Society for the Promotion of Science (JSPS) Grants Numbers 
JP21H01005, JP22H01140 and JP24K00545.

\appendix
\section{Algebraic structure of the effective Hamiltonian}
\label{sec:algebraic-structure}

We analyze an algebraic structure of
the eigenspace of the effective Hamiltonian $H_{\rm eff}$ in Eq.~\eqref{eq:eff-Hamiltonian}
in the simple case $v_{m\alpha}=v$, $\epsilon_{{\rm d}\alpha}=\epsilon_{\rm d}$,
$v^{\prime}=v^{\prime\ast}$ and $U_{\alpha}=U$ for $m=1, 2$ and $\alpha=1, 2$.
In terms of the creation- and annihilation-operators of electrons on the quantum dots, 
we introduce the operators associated to spin degrees of freedom as
\begin{eqnarray}
\label{eq:spin-su(2)}
 S_{z}=\frac{1}{2}\sum_{\alpha}(n_{\alpha\uparrow}-n_{\alpha\downarrow}),\quad
 S_{+}=\sum_{\alpha}d^{\dagger}_{\alpha\uparrow}d_{\alpha\downarrow},\quad
 S_{-}=\sum_{\alpha}d^{\dagger}_{\alpha\downarrow}d_{\alpha\uparrow}
\end{eqnarray}
and those associated to charge degrees of freedom as
\begin{eqnarray}
\label{eq:charge-su(2)}
\fl
 \eta_{z}=\frac{1}{2}\sum_{\alpha}(1-n_{\alpha\uparrow}-n_{\alpha\downarrow}),\quad
 \eta_{+}=\sum_{\alpha}(-1)^{\overline{\alpha}}d_{\alpha\downarrow}d_{\alpha\uparrow},\quad
 \eta_{-}=\sum_{\alpha}(-1)^{\overline{\alpha}}
 d^{\dagger}_{\alpha\uparrow}d^{\dagger}_{\alpha\downarrow}.
\end{eqnarray}
The operators $S_{z}$ and $\eta_{z}$ characterize the numbers of 
up-spins and down-spins of electrons on the two quantum dots.
In fact, the state $|N, M\rangle$ with $N$ electrons and $M$ down-spins
for $0\leq M\leq N\leq 2$
is a joint eigenstate of the operators $S_{z}$ and $\eta_{z}$ as in
\begin{eqnarray}
 S_{z}|N, M\rangle=\frac{1}{2}(N-2M)|N, M\rangle,\quad
 \eta_{z}|N, M\rangle=\frac{1}{2}(2-N)|N, M\rangle.
\end{eqnarray}
The operators $S_{\pm}$ flip the spin of an electron
and the operators $\eta_{\pm}$ create or annihilate a pair of electrons with opposite spins.
Each set of the operators in Eqs.~\eqref{eq:spin-su(2)} and \eqref{eq:charge-su(2)} 
gives a representation of the Lie algebra $su(2)$:
\begin{eqnarray}
&[S_{z},S_{\pm}]=\pm S_{\pm},\quad [S_{+},S_{-}]=2S_{z},
 \nn\\
&[\eta_{z},\eta_{\pm}]=\pm \eta_{\pm},\quad [\eta_{+},\eta_{-}]=2\eta_{z}.
\end{eqnarray}
The representation $\{S_{z}, S_{\pm}\}$ is referred to as spin-$su(2)$
and $\{\eta_{z}, \eta_{\pm}\}$ is referred to as charge-$su(2)$.
Since all the operators in the set $\{S_{z}, S_{\pm}\}$ are commutative
with those in the set $\{\eta_{z}, \eta_{\pm}\}$, the set $\{S_{z}, S_{\pm}, \eta_{z}, \eta_{\pm}\}$
gives a representation of the semisimple Lie algebra $su(2)\oplus su(2)\simeq so(4)$.
We remark that the operators in Eqs.~\eqref{eq:spin-su(2)} and \eqref{eq:charge-su(2)}
were first introduced in order to elucidate the $SO(4)$ symmetry 
of the one-dimensional Hubbard model~\cite{Yang_89PRL,Yang-Zhang_90MPL}.

Let us consider irreducible highest-weight representations of the algebra $so(4)$.
We introduce the Casimir operators for each representation in Eqs.~\eqref{eq:spin-su(2)} and \eqref{eq:charge-su(2)} as
\begin{eqnarray}
 \Vec{S}^{2}=S_{z}^{2}+\frac{1}{2}(S_{+}S_{-}+S_{-}S_{+}),\quad
 \Vec{\eta}^{2}=\eta_{z}^{2}+\frac{1}{2}(\eta_{+}\eta_{-}+\eta_{-}\eta_{+}).
\end{eqnarray}
Let $|S, \eta\rangle\!\rangle$ be a highest-weight state in the representation space 
that satisfies the relations
\begin{eqnarray}
&S_{+}|S, \eta\rangle\!\rangle=\eta_{+}|S, \eta\rangle\!\rangle=0,
 \nn\\
&\Vec{\it S}^{2}|S, \eta\rangle\!\rangle=S(S+1)|S, \eta\rangle\!\rangle,\quad
 \Vec{\eta}^{2}|S, \eta\rangle\!\rangle=\eta(\eta+1)|S, \eta\rangle\!\rangle
\end{eqnarray}
with non-negative real numbers $S$ and $\eta$.
Successive action of the operators in Eqs.~\eqref{eq:spin-su(2)} and \eqref{eq:charge-su(2)} 
on the highest-weight state $|S, \eta\rangle\!\rangle$ generates a highest-weight representation space.
If both $2S+1$ and $2\eta+1$ are positive integers, the highest-weight representation 
gives a $(2S+1)(2\eta+1)$-dimensional irreducible representation of the algebra $so(4)$.
The states $(S_{-})^{n}(\eta_{-})^{m}|S, \eta\rangle\!\rangle$ 
for $n=0, 1, \ldots, 2S$ and $m=0, 1, \ldots, 2\eta$ constitute a basis set 
of the irreducible-representation space.

\begin{figure}[t]
\begin{center}
{
\begin{picture}(280,240)(0,0)
\put(0,-5){\includegraphics[width=280pt]{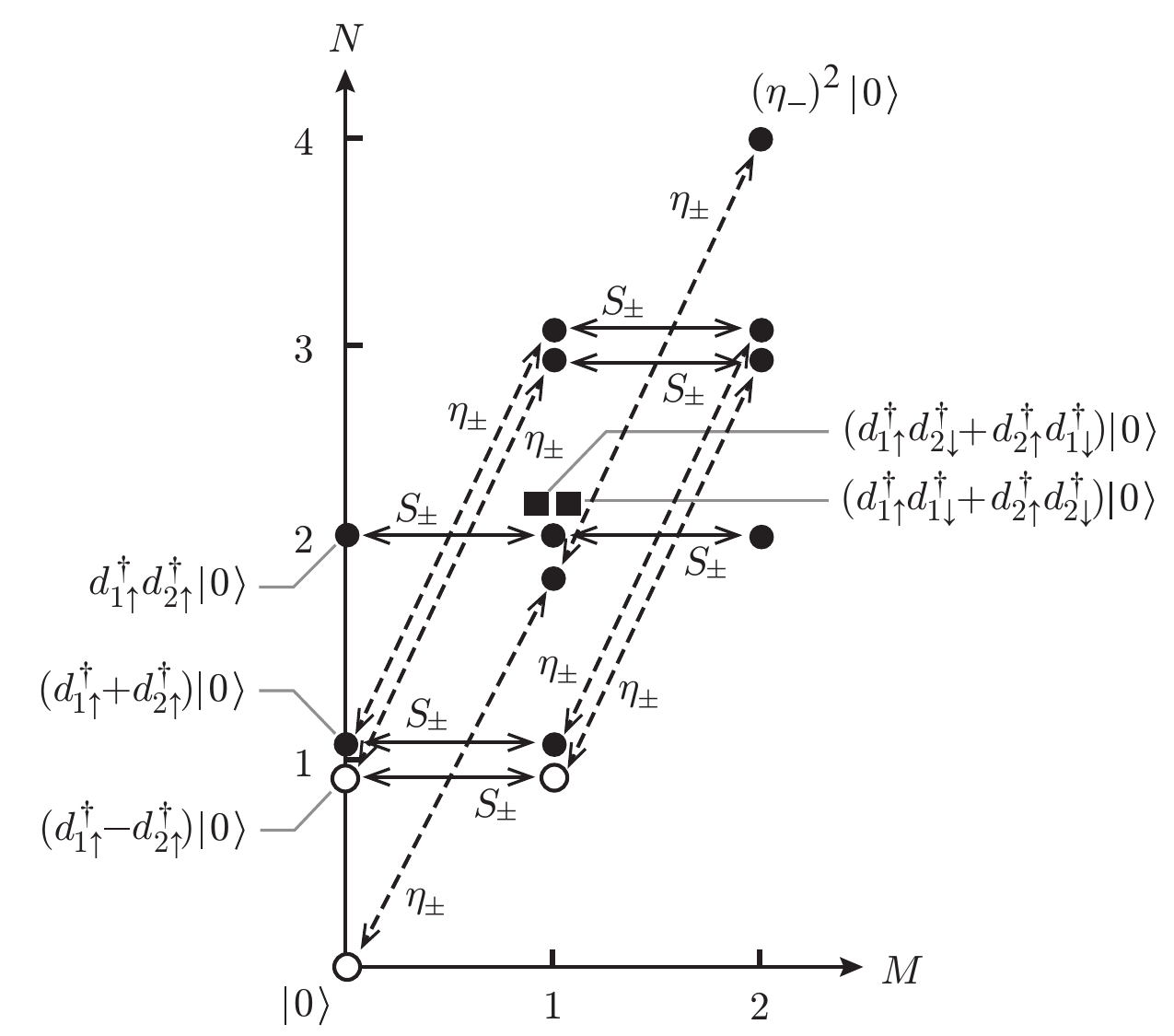}}
\end{picture}
}
\end{center}
\caption{\label{fig:eigenspace} 
$so(4)$-algebraic structure of the eigenspace of the effective Hamiltonian $H_{\rm eff}$.
The vertical axis represents the number of electrons $N$
and the horizontal axis represents that of down-spins $M$.
Each white circle corresponds to a bound state, 
each black circle corresponds to a resonant eigenstate that is a highest or a descendant state of $so(4)$,
and each black square corresponds to a resonant eigenstate that is a superposition of two singlet states of $so(4)$.
The solid double-headed arrows represent the action of $S_{\pm}$
and the dashed ones represent the action of $\eta_{\pm}$.}
\end{figure}

The 16-dimensional electron-state space on the two quantum dots that
the effective Hamiltonian $H_{\rm eff}$ in Eq.~\eqref{eq:eff-Hamiltonian} acts on is decomposed 
into the direct sum of irreducible representations of the algebra $so(4)$.
We find six highest-weight states
$|0\rangle$, $(d^{\dagger}_{1\uparrow}\pm d^{\dagger}_{2\uparrow})|0\rangle$, 
$d^{\dagger}_{1\uparrow}d^{\dagger}_{2\uparrow}|0\rangle$, 
$(d^{\dagger}_{1\uparrow}d^{\dagger}_{1\downarrow}+d^{\dagger}_{2\uparrow}d^{\dagger}_{2\downarrow})|0\rangle$
and $(d^{\dagger}_{1\uparrow}d^{\dagger}_{2\downarrow}+d^{\dagger}_{2\uparrow}d^{\dagger}_{1\downarrow})|0\rangle$,
as is depicted in Figure~\ref{fig:eigenspace}.
The action of the operators in Eqs.~\eqref{eq:spin-su(2)} and \eqref{eq:charge-su(2)} to
each highest-weight state, which is indicated by the arrows on Figure~\ref{fig:eigenspace},
generates the following irreducible representations:
\begin{itemize}
\item[(I)] The vacuum state $|0\rangle$, which is located at $(N, M)=(0, 0)$ on Figure~\ref{fig:eigenspace},
is a highest-weight state $|S, \eta\rangle\!\rangle$ with $(S, \eta)=(0, 1)$
and descendant states in the irreducible highest-weight representation are given by
\begin{eqnarray}
  \eta_{-}|0\rangle=(d^{\dagger}_{1\uparrow}d^{\dagger}_{1\downarrow}-d^{\dagger}_{2\uparrow}d^{\dagger}_{2\downarrow})|0\rangle,\quad
 (\eta_{-})^{2}|0\rangle=2d^{\dagger}_{1\uparrow}d^{\dagger}_{2\uparrow}d^{\dagger}_{1\downarrow}d^{\dagger}_{2\downarrow}|0\rangle,
\end{eqnarray}
which form a singlet state for the spin-$su(2)$ and triplet states for the charge-$su(2)$.
\item[(II)] The state $d^{\dagger}_{1\uparrow}d^{\dagger}_{2\uparrow}|0\rangle$, 
which is located at $(N, M)=(2, 0)$ on Figure~\ref{fig:eigenspace},
is a highest-weight state $|S, \eta\rangle\!\rangle$ with $(S, \eta)=(1, 0)$
and descendant states in the irreducible highest-weight representation are given by
\begin{eqnarray}
\fl
  S_{-}d^{\dagger}_{1\uparrow}d^{\dagger}_{2\uparrow}|0\rangle
 =(d^{\dagger}_{1\uparrow}d^{\dagger}_{2\downarrow}-d^{\dagger}_{2\uparrow}d^{\dagger}_{1\downarrow})|0\rangle,\quad
 (S_{-})^{2}d^{\dagger}_{1\uparrow}d^{\dagger}_{2\uparrow}|0\rangle
 =2d^{\dagger}_{1\downarrow}d^{\dagger}_{2\downarrow}|0\rangle,
\end{eqnarray}
which form triplet states for the spin-$su(2)$ and a singlet state for the charge-$su(2)$.
\item[(III)]  The state 
$(d^{\dagger}_{1\uparrow}d^{\dagger}_{2\downarrow}
+d^{\dagger}_{2\uparrow}d^{\dagger}_{1\downarrow})|0\rangle$, 
which is located at $(N, M)=(2, 1)$ on Figure~\ref{fig:eigenspace},
is a highest-weight state $|S, \eta\rangle\!\rangle$ with $(S, \eta)=(0,0)$.
This state is a singlet state for both the spin-$su(2)$ and the charge-$su(2)$.
\item[(IV)] The state 
$(d^{\dagger}_{1\uparrow}d^{\dagger}_{1\downarrow}
+d^{\dagger}_{2\uparrow}d^{\dagger}_{2\downarrow})|0\rangle$, 
which is also located at $(N, M)=(2, 1)$ on Figure~\ref{fig:eigenspace},
is also a highest-weight state $|S, \eta\rangle\!\rangle$ with $(S, \eta)=(0,0)$.
This state is again a singlet state for both the spin-$su(2)$ and the charge-$su(2)$.
\item[(V)] The states $(d^{\dagger}_{1\uparrow}\pm d^{\dagger}_{2\uparrow})|0\rangle$, 
which are both located at $(N, M)=(1, 0)$ on Figure~\ref{fig:eigenspace},
are highest-weight states $|S, \eta\rangle\!\rangle$ with $(S, \eta)=(1/2,1/2)$
and descendant states in the irreducible highest-weight representation are given by
\begin{eqnarray}
&S_{-}(d^{\dagger}_{1\uparrow}\pm d^{\dagger}_{2\uparrow})|0\rangle
 =(d^{\dagger}_{1\downarrow}\pm d^{\dagger}_{2\downarrow})|0\rangle,
  \nn\\
&\eta_{-}(d^{\dagger}_{1\uparrow}\pm d^{\dagger}_{2\uparrow})|0\rangle
 =\mp d^{\dagger}_{1\uparrow}d^{\dagger}_{2\uparrow}
  (d^{\dagger}_{1\downarrow}\pm d^{\dagger}_{2\downarrow})|0\rangle,
 \nn\\
&S_{-}\eta_{-}(d^{\dagger}_{1\uparrow}\pm d^{\dagger}_{2\uparrow})|0\rangle
 =\pm(d^{\dagger}_{1\uparrow}\pm d^{\dagger}_{2\uparrow})
 d^{\dagger}_{1\downarrow}d^{\dagger}_{2\downarrow}|0\rangle,
\end{eqnarray}
which form doublet states for both the spin-$su(2)$ and the charge-$su(2)$.
\end{itemize}
These states form a basis of the 16-dimensional electron-state space on the two quantum dots.

Let us analyze the relation between the states on Figure~\ref{fig:eigenspace}
and the eigenstates of the effective Hamiltonian $H_{\rm eff}$ in Eq.~\eqref{eq:eff-Hamiltonian}.
First, we find that the descendant state $S_{-}d^{\dagger}_{1\uparrow}d^{\dagger}_{2\uparrow}|0\rangle
 =(d^{\dagger}_{1\uparrow}d^{\dagger}_{2\downarrow}-d^{\dagger}_{2\uparrow}d^{\dagger}_{1\downarrow})|0\rangle$ 
in the spin-triplet states in the case (II) is the eigenstate with the eigenfunction
$f^{\rm (II)}_{\alpha\beta,\uparrow\downarrow}(t)$ in Eqs.~\eqref{eq:time-evolving-resonant-state_so(4)}
and the eigenvalue $E^{(2)}_{{\rm R}, 2}$.
Due to the spin-$su(2)$ symmetry of the effective Hamiltonian $H_{\rm eff}$, which 
is shown by the commutativity 
\begin{eqnarray}
\label{eq:spin-su(2)_H}
 [H_{\rm eff}, S_{z}]=[H_{\rm eff}, S_{\pm}]=0,
\end{eqnarray}
the other states $d^{\dagger}_{1\uparrow}d^{\dagger}_{2\uparrow}|0\rangle$ and 
$(S_{-})^{2}d^{\dagger}_{1\uparrow}d^{\dagger}_{2\uparrow}|0\rangle
=2d^{\dagger}_{1\downarrow}d^{\dagger}_{2\downarrow}|0\rangle$ that form the spin-triplet states
are also eigenstates of $H_{\rm eff}$ with the same eigenvalue $E^{(2)}_{{\rm R}, 2}$.
This shows that
the time evolution of the initial state $S_{-}d^{\dagger}_{1\uparrow}d^{\dagger}_{2\uparrow}|0\rangle$
is equivalent to that in the spinless two-electron case discussed in Ref.~\cite{Nishino-Hatano_24JPA}.

Second, we find that the descendant state $\eta_{-}|0\rangle
=(d^{\dagger}_{1\uparrow}d^{\dagger}_{1\downarrow}-d^{\dagger}_{2\uparrow}d^{\dagger}_{2\downarrow})|0\rangle$ 
in the charge-triplet states in the case (I) is the eigenstate with the eigenfunction
$f^{\rm (I)}_{\alpha\beta,\uparrow\downarrow}(t)$ in Eqs.~\eqref{eq:time-evolving-resonant-state_so(4)} 
and the eigenvalue $E^{(2)}_{{\rm R}, 1}$.
In contrast to the spin-$su(2)$ symmetry, 
the effective Hamiltonian $H_{\rm eff}$ does not have the charge-$su(2)$ symmetry;
the effective Hamiltonian $H_{\rm eff}$ is commutative only with the operator $\eta_{z}$,
but not with $\eta_{\pm}$ as in
\begin{eqnarray}
\label{eq:charge-su(2)_H}
&[H_{\rm eff}, \eta_{+}]
 =-\big(2(\epsilon_{{\rm d}}-\ii\Gamma)+U+2U^{\prime}N\big)\eta_{+},
  \nn\\
&[H_{\rm eff}, \eta_{-}]
 =\eta_{-}\big(2(\epsilon_{{\rm d}}-\ii\Gamma)+U+2U^{\prime}N\big),
\end{eqnarray}
where $N=\sum_{\alpha, \sigma}n_{\alpha\sigma}$.
The second relation shows that the states $|0\rangle$, $\eta^{-}|0\rangle$ and $(\eta^{-})^{2}|0\rangle$
in the charge-triplet states in the case (I) are the eigenstates of the effective Hamiltonian $H_{\rm eff}$ 
with different eigenvalues as in
\begin{eqnarray}
\fl
&H_{\rm eff}|0\rangle=0|0\rangle,
  \nn\\
\fl
&H_{\rm eff}\,\eta_{-}|0\rangle
  =\eta_{-}\big(H_{\rm eff}+2(\epsilon_{{\rm d}}-\ii\Gamma)+U+2U^{\prime}N\big)|0\rangle
  =E^{(2)}_{{\rm R}, 1}\eta_{-}|0\rangle,
  \nn\\
\fl
&H_{\rm eff}\,(\eta_{-})^{2}|0\rangle
  =\eta_{-}\big(H_{\rm eff}+2(\epsilon_{{\rm d}}-\ii\Gamma)+U+2U^{\prime}N\big)\eta_{-}|0\rangle
  =(2E^{(2)}_{{\rm R}, 1}+4U^{\prime})(\eta_{-})^{2}|0\rangle.
\end{eqnarray}
We remark that the Hamiltonian $H_{\rm eff}$ has the charge-$su(2)$ symmetry
only in the case of $2\epsilon_{{\rm d}}+U=U^{\prime}=0$ and $\Gamma=0$,
which is equivalent to the two-site Hubbard model with a particle-hole symmetry. 

Third, we find that
the one-dimensional representation space spanned by the singlet state 
$(d^{\dagger}_{1\uparrow}d^{\dagger}_{2\downarrow}+d^{\dagger}_{2\uparrow}d^{\dagger}_{1\downarrow})|0\rangle$ 
in the case (III) as well as the one by 
$(d^{\dagger}_{1\uparrow}d^{\dagger}_{1\downarrow}+d^{\dagger}_{2\uparrow}d^{\dagger}_{2\downarrow})|0\rangle$ 
in the case (IV)
is not invariant under the action of the effective Hamiltonian $H_{\rm eff}$.
In fact, as we have already seen in Eq.~\eqref{eq:eff-Hamiltonian_2-elec_3} of Section~\ref{sec:resonance-energy},
the effective Hamiltonian $H_{\rm eff}$ becomes a 2$\times$2 block on the space
spanned by these two singlet states.
In order to obtain singlet states that give a representation space 
invariant under the action of the effective Hamiltonian $H_{\rm eff}$,
we need to take a superposition of the two singlet states
$(d^{\dagger}_{1\uparrow}d^{\dagger}_{2\downarrow}+d^{\dagger}_{2\uparrow}d^{\dagger}_{1\downarrow})|0\rangle$ 
and $(d^{\dagger}_{1\uparrow}d^{\dagger}_{1\downarrow}+d^{\dagger}_{2\uparrow}d^{\dagger}_{2\downarrow})|0\rangle$.

Fourth, the highest state $(d^{\dagger}_{1\uparrow}\pm d^{\dagger}_{2\uparrow})|0\rangle$ in the case (V)
is the one-electron eigenstate of the effective Hamiltonian $H_{\rm eff}$
with the eigenvalue $E^{(1)}_{{\rm R}\mp}$ in Eq.~\eqref{eq:1-resonance-energy_2}, 
which we showed in the previous work~\cite{Nishino-Hatano_24JPA}.
All the descendant states are eigenstates of $H_{\rm eff}$, which are shown as
\begin{eqnarray}
\fl
&H_{\rm eff}(d^{\dagger}_{1\uparrow}\pm d^{\dagger}_{2\uparrow})|0\rangle
 =E^{(1)}_{{\rm R}\mp}(d^{\dagger}_{1\uparrow}\pm d^{\dagger}_{2\uparrow})|0\rangle,
 \nn\\
\fl
&H_{\rm eff}S_{-}(d^{\dagger}_{1\uparrow}\pm d^{\dagger}_{2\uparrow})|0\rangle
 =E^{(1)}_{{\rm R}\mp}S_{-}(d^{\dagger}_{1\uparrow}\pm d^{\dagger}_{2\uparrow})|0\rangle,
  \nn\\
\fl
&H_{\rm eff}\,\eta_{-}(d^{\dagger}_{1\uparrow}\pm d^{\dagger}_{2\uparrow})|0\rangle
  =\big(E^{(1)}_{{\rm R}\mp}+2(\epsilon_{{\rm d}}-\ii\Gamma)+U+2U^{\prime}\big)
   \eta_{-}(d^{\dagger}_{1\uparrow}\pm d^{\dagger}_{2\uparrow})|0\rangle,
  \nn\\
\fl
&H_{\rm eff}\,S_{-}\eta_{-}(d^{\dagger}_{1\uparrow}\pm d^{\dagger}_{2\uparrow})|0\rangle
  =\big(E^{(1)}_{{\rm R}\mp}+2(\epsilon_{{\rm d}}-\ii\Gamma)+U+2U^{\prime}\big)
   S_{-}\eta_{-}(d^{\dagger}_{1\uparrow}\pm d^{\dagger}_{2\uparrow})|0\rangle.
\end{eqnarray}

\section*{References}

\begin{thebibliography}{10}
\bibitem{Gamow_28ZPhysA}
Gamow V G
1928 {\it Z.~Phys.~A} {\bf 51} 204
\bibitem{Siegert_39PR}
Siegert A J F 
1939 {\it Phys.~Rev.} {\bf 56} 750
\bibitem{Zeldovich_61JETP}
Zel'dovich Y B
1961 {\it Sov.~Phys.~JETP.} {\bf 12} 542
\bibitem{Hokkyo_65PTP}
Hokkyo N 
1965 {\it Prog.~Theor.~Phys.} {\bf 33} 1116
\bibitem{Berggren_68NPA}
Berggren T
1968 {\it Nucl.~Phys.~A} {\bf 109} 265
\bibitem{Aguilar-Combes_71CMP}
Aguilar J and Combes J M
1971 {\it Commun.~Math.~Phys.} {\bf 22} 269
\bibitem{Balslev-Combes_71CMP}
Balslev E and Combes J M
1971 {\it Commun.~Math.~Phys.} {\bf 22} 280
\bibitem{Simon_72CMP}
Simon B
1972 {\it Commun.~Math.~Phys.} {\bf 27} 1
\bibitem{Reinhardt_82AnnuRevPhysChem}
Reinhardt W P 
1982 {\it Annu.~Rev.~Phys.~Chem.} {\bf 33} 223
\bibitem{Moiseyev_98PR}
Moiseyev N
1998 {\it Phys.~Rep.} {\bf 302} 211
\bibitem{Moiseyev_11NHQP}
Moiseyev N 
2011 {\it Non-Hermitian Quantum Mechanics}, (Cambridge University Press)
\bibitem{Romo_68NPA}
Romo W 
1968 {\it Nucl. Phys.} A {\bf 116} 618
\bibitem{GarciaCalderon-Peierls_76NPA}
Garc\'{\i}a-Calder\'on G and Peierls R 
1976 {\it Nucl.~Phys.~A} {\bf 265} 443
\bibitem{Berrondo-GarciaCalderon_77LNC}
Berrondo M and Garc\'{\i}a-Calder\'on G 
1977 {\it Lett.~Nuovo Cimento} {\bf 20} 34
\bibitem{Lind_93PRA}
Lind P 
1993 {\it Phys. Rev. C} {\bf 47} 1903
\bibitem{More_71PRA}
More R M 
1971 {\it Phys.~Rev.~A} {\bf 4} 1782
\bibitem{More-Gerjuoy_73PR}
More R M and Gerjuoy E
1973 {\it Phys.~Rev.} {\bf 7} 1288
\bibitem{GarciaCalderon_76NPA}
Garc\'{\i}a-Calder\'on G
1976 {\it Nucl.~Phys.~A} {\bf 261} 130
\bibitem{GarciaCalderon_82LNC}
Garc\'{\i}a-Calder\'on G
1982 {\it Lett.~Nuovo Cimento} {\bf 33} 253
\bibitem{GarciaCalderon-Rubio_86NPA}
Garc\'{\i}a-Calder\'on G and Rubio A
1986 {\it Nucl.~Phys.~A} {\bf 458} 560
\bibitem{GarciaCalderon-Romo-Rubio_93PRB}
Garc\'{\i}a-Calder\'on G, Romo R and Rubio A
1993 {\it Phys.~Rev.~B} {\bf 47} 9572
\bibitem{Romo-GarciaCalderon_94PRB}
Romo R and Garc\'{\i}a-Calder\'on G 
1994 {\it Phys.~Rev.~B} {\bf 49} 14016
\bibitem{GarciaCalderon-Romo-Rubio_97PRB}
Garc\'{\i}a-Calder\'on G, Romo R and Rubio A
1997 {\it Phys.~Rev.~B} {\bf 56} 4845
\bibitem{Hatano-Sasada-Nakamura-Petrosky_08PTP}
Hatano N, Sasada K, Nakamura H and Petrosky T
2008 {\it Prog. Theor. Phys.} {\bf 119} 187
\bibitem{Hatano-Ordonez_14JMP}
Hatano H and Ordonez G
2014 {\it J.~Math.~Phys.} {\bf 55} 122106
\bibitem{Ordonez-Hatano_17JPA}
Ordonez G and Hatano N 
2017 {\it J.~Phys.~A:~Math.~Theor.} {\bf 50} 405304
\bibitem{Ordonez-Hatano_17Chaos}
Ordonez G and Hatano N 
2017 {\it Chaos.} {\bf 27} 104608
\bibitem{Hatano_21JPCS}
Hatano H
2021 {\it J.~Phys.:~Conf.~Ser.} {\bf 2038} 012013
\bibitem{Feshbach_58ARNS} 
Feshbach H
1958 {\it Annual Review of Nuclear Science} {\bf 8} 49
\bibitem{Feshbach_58AP}
Feshbach H 
1958 {\it Ann.~Phys.~(NY)} {\bf 5} 357 
\bibitem{Feshbach_62AP}
Feshbach H
1962 {\it Ann.~Phys.~(NY)} {\bf 19} 287
\bibitem{Nishino-Hatano_24JPA}
Nishino A and Hatano N
2024 {\it J.~Phys.~A:~Math.~Theor.} {\bf 57} 245302
\bibitem{Hatano-Kawamoto-Feinberg_09PJP}
Hatano N, Kawamoto T and Feinberg J
2009 {\it Pramana J.~Phys.} {\bf 73} 553 
\bibitem{Nishino-Imamura-Hatano_12JPC}
Nishino A, Hatano N and Ordonez G
2012 {\it J.~Phys.:~Conf.~Ser.} {\bf 343} 012087
\bibitem{Nishino-Hatano-Ordonez_16JPC}
Nishino A, Hatano N and Ordonez G
2016 {\it J.~Phys.:~Conf.~Ser.} {\bf 670} 012038
\bibitem{Petrosky-Tasaki-Prigogine_91PhysicaA}
Petrosky T, Tasaki S and Prigogine I 
1991 {\it Physica A} {\bf 170} 306
\bibitem{Khalfin_68PZETF}
Khalfin L A, 
1968 {\it Pis'ma Zh.~Eksp.~Teor.~Fiz.} {\bf 8}, 106 [1968 JETP Letters {\bf 8} 65].
\bibitem{Chiu-Sudarshan-Misra_77PRD}
Chiu C B, Sudarshan E C G and Misra B
1977 {\it Phys.~Rev.~D} {\bf 16} 520
\bibitem{Petrosky-Ordonez-Prigogine_01PRA}
Petrosky T, Ordonez G and Prigogine I
2001 {\it Phys.~Rev.~A} {\bf 64} 062101
\bibitem{Garmon-Petrosky-Simine-Segal_13FortschrPhys}
Garmon A, Petrosky T, Simine L and Segal D
2013 {\it Fortschr.~Phys.} {\bf 61} 261
\bibitem{Chakraborty-Sensarma_18PRB}
Chakraborty A and Sensarma R
2018 {\it Phys.~Rev.~B} {\bf 97} 104306
\bibitem{Garmon-Noba-Ordonez-Segal_19PRD}
Garmon S, Noba K, Ordonez G and Segal D
2019 {\it Phys.~Rev.~A} {\bf 99} 010102
\bibitem{Taira-Hatano-Nishino_2024preprint}
Taira T, Hatano N and Nishino A 
{\it preprint} (arXiv:2406.17436).
\bibitem{Itano-Heizen-Bollinger-Wineland_90PRA}
Itano W M, Heinzen D J, Bollinger J J and Wineland D J
1990 {\it Phys.~Rev.~A} {\bf 41} 2295
\bibitem{Wilkinson_97Nature}
Wilkinson S R et al.
1997 {\it Nature} {\bf 387} 575
\bibitem{Fischer-GutierrezMedina-Raizen_01PRL}
Fischer M C, Guti\'errez-Medina B and Raizen M G
2001 {\it Phys.~Rev.~Lett.} {\bf 87} 040402
\bibitem{Fujisawa_06Science}
Fujisawa T, Hayashi T, Tomita R and Hirayama Y
2006 {\it Science} {\bf 312} 1634
\bibitem{Alexander-Anderson_64JPA}
Alexander S and Anderson P W 
1964 {\it Phys.~Rev.~A} {\bf 133} 1594
\bibitem{Bender-Boettcher_98PRL}
Bender C M and Boettcher S
1998 {\it Phys.~Rev.~Lett.} {\bf 80} 5243 
\bibitem{Sasada-Hatano_08JPSJ}
Sasada K and Hatano N
2008 {\it J.~Phys.~Soc.~Jpn.} {\bf 77} 025003
\bibitem{Yang_89PRL}
Yang C N, 
1989 {\it Phys.~Rev.~Lett.} {\bf 63} 2144
\bibitem{Yang-Zhang_90MPL}
Yang C N and Zhang S C
1990 {\it Mod.~Phys.~Lett.~B} {\bf 4} 759
\end{thebibliography}

\end{document}